\documentclass{article}
\usepackage{fullpage}
\usepackage[usenames]{color}
\usepackage[colorlinks = true]{hyperref}
\usepackage[english]{babel}
\usepackage{url}
\usepackage[caption=false,font=normalsize,labelfont=sf,textfont=sf]{subfig}
\usepackage[title]{appendix}
\usepackage{tikz}
\usetikzlibrary{calc,patterns,decorations.pathmorphing,decorations.markings}

\usepackage{bm,bbm,amsmath,amssymb,multicol,color}
\usepackage{cleveref,cite}

\newcommand  \stack[2]   {\overset{\text{#1}}{#2}}
\usepackage{algorithm,pifont}

\newcommand{\MAT}{\left[ \begin{array}}  
\newcommand{\mat}{\end{array} \right]}
\newtheorem{Definition}{Definition}[section]

\newtheorem{Lemma}{Lemma}[section]
\newtheorem{Theorem}{Theorem}[section]

\newtheorem{Q}{Question}[section]

\def \A {\mathbf{A}}
\def \AA {\mathcal{A}}

\def \e {\bm{e}}
\def \E {\mathbf{E}}
\def \EE{\mathcal{E}}
\def \EEE{\mathbb{E}}

\def \F {\mathbf{F}}

\def \G {\mathbf{G}}
\def \h {\bm{h}}

\def \I {\mathbf{I}}
\def \J {\mathbf{J}}

\def \NN {\mathcal{N}}

\def \PPP {\mathbb{P}}
\def \q {\bm{q}}
\def \Q {\mathbf{Q}}

\def \R {\mathbf{R}}

\def \RRR {\mathbb{R}}
\def \s{\bm{s}}
\def \S {\mathbf{S}}

\def \U {\mathbf{U}}

\def \V {\mathbf{V}}

\def \x {\bm{x}}

\def \X {\mathbf{X}}
\def \XX {\mathcal{X}}

\def \Xh {\widehat{\mathbf{X}}}
\def \Xs {\mathbf{X}^\star}

\def \y {\bm{y}}
\def \Y {\mathbf{Y}}
\def \z {\bm{z}}
\def \Z {\mathbf{Z}}

\def \sumt{\sum_{t=1}^T}
\def \summ{\sum_{m=1}^M}

\def \bSigma {\boldsymbol{\Sigma}}
\def \bGamma {\boldsymbol{\Gamma}}

\def \one {\mathbf{1}}

\begin{document}

 
\title{Recovery Guarantees for Time-varying Pairwise Comparison Matrices with Non-transitivity}

\author{Shuang Li and Michael B. Wakin\thanks{SL is with the Department of Mathematics, University of California, Los Angeles. Email: shuangli@math.ucla.edu. MBW is with the Department of Electrical Engineering, Colorado School of Mines. Email: mwakin@mines.edu.}}

\maketitle

\begin{abstract}
Pairwise comparison matrices have received substantial attention in a variety of applications, especially in rank aggregation, the task of flattening items into a one-dimensional (and thus transitive) ranking. However, non-transitive preference cycles can arise in practice due to the fact that making a decision often requires a complex evaluation of multiple factors. In some applications, it may be important to identify and preserve information about the inherent non-transitivity, either in the pairwise comparison data itself or in the latent feature space. In this work, we develop structured models for non-transitive pairwise comparison matrices that can be exploited to recover such matrices from incomplete noisy data and thus allow the detection of non-transitivity. Considering that individuals' tastes and items' latent features may change over time, we formulate time-varying pairwise comparison matrix recovery as a dynamic skew-symmetric matrix recovery problem by modeling changes in the low-rank factors of the pairwise comparison matrix. We provide theoretical guarantees for the recovery and numerically test the proposed theory with both synthetic and real-world data.

\end{abstract}


\section{Introduction}
Pairwise comparison matrices are data structures that can arise in a number of applications including recommendation engines, economic exchanges, elections, and psychology~\cite{kendall1940method,david1987ranking,ma2006matrix,hochbaum2010separation,jiang2011statistical,gleich2011rank}. In practice, pairwise comparisons can be made either directly (e.g., by observing the outcome of a competition between items $i$ and $j$) or indirectly (e.g., by aggregating a collection of voter/item ratings such as the Netflix data set~\cite{gleich2011rank}). Thus, real-world observations of such matrices may naturally be incomplete, noisy, or involve some degree of randomness. Given the available pairwise comparisons, denoising the entries or inferring the missing entries of the matrix can be valuable for making better decisions and recommendations. Fortunately, such matrices can possess an intrinsic structure that makes such inference, denoising, and recovery possible.

To date, pairwise comparison matrices have received substantial attention in solving a problem known as {\em rank aggregation}~\cite{david1987ranking,hochbaum2010separation,jiang2011statistical,gleich2011rank,borda1784memoire,kemeny1959mathematics,freund2003efficient,d2021ranking}, where one seeks an ordered ranking of $N$ items in the list that best agrees with available votes or ratings. In fact, there is a natural relationship between the rank aggregation problem and a structured model for a quantitative pairwise comparison matrix $\X\in\RRR^{N\times N}$: supposing that each item $i$ possesses an intrinsic value $s_i$ and that $\X$ is populated according to the rule $X_{ij} := s_i-s_j$, it follows that
\begin{equation}
	\X = \s\one^\top - \one\s^\top,
    \label{eq:rank2trans}
\end{equation}
where $\s = \begin{bmatrix} s_1 & s_2 & \cdots & s_N \end{bmatrix}^\top$ is a vector containing the value parameters and $\one\in\RRR^N$ is a vector containing all ones. Then $\X$ has rank equal to only 2, and it is skew-symmetric (i.e., $\X = -\X^\top$). Using this model, Gleich and Lim~\cite{gleich2011rank} propose an algorithm for recovering pairwise comparison matrices from incomplete and inaccurate measurements (and from those, recovering the values $s_i$). Unfortunately, since it is based on a latent one-dimensional ordering of the values $s_i$, this model assumes and enforces that the pairwise comparisons be {\em transitive}. \textcolor{black}{In particular, \eqref{eq:rank2trans} implies both cardinal transitivity ($\X(i,j) = \X(i,k) + \X(k,j)$ for all $i,j,k$) and ordinal transitivity ($\X(i,k) > 0$ and $\X(k,j) > 0$ $\Rightarrow$ $X(i,j) > 0$) in pairwise comparisons.}

In many rank aggregation techniques, {\em non-transitivities} in $\X$ are treated as nuisances that must be overcome in order to find the ``most consistent" global ranking of the items~\cite{hochbaum2010separation,gleich2011rank,kemeny1959mathematics,saaty1984inconsistency}. However, in many real-world settings pairwise relationships can easily be non-transitive~\cite{jiang2011statistical,marquis1785essai,chen2016modeling,makhijani2018social}, especially if the comparison of two items depends on multiple latent factors rather than a single scalar quantity. In such settings, flattening a pairwise comparison matrix into a one-dimensional ranking can destroy important information about the relationships among the items. Therefore, it is important to identify and preserve information about the inherent non-transitivity, either in the pairwise comparison data itself or in the latent feature space.

In work~\cite{yang2015modeling}, the authors extend the above model to account for non-transitive preferences. Doing so requires using a higher-dimensional model for the latent parameter space, thus capturing the fact that many preferences rely on multiple underlying factors. Suppose each item is characterized by $2r$ latent properties, which altogether can be represented in matrices $\S \in \RRR^{N\times r}$ and $\Q \in \RRR^{N\times r}$. Then the proposed model $\X = \S\Q^\top - \Q\S^\top$ captures the possibility that properties of item $j$ might inhibit properties of item $i$, and vice versa. (See~\cite{yang2015modeling} for further details; see also \cite{chen2016modeling} for an equivalent Blade-Chest model with $r=1$.) The resulting matrix $\X$ will again be skew-symmetric and will have rank at most $2r$, thus preserving the low-rank property when $r$ is small compared to $N$. However, and importantly, \textcolor{black}{cardinal and ordinal} non-transitive preferences can exist in $\X$. In~\cite{yang2015modeling}, these facts are used to reconstruct such non-transitive matrices when only certain entries are observed.

Meanwhile, individuals' tastes can change over time; candidates in an election adapt their platforms; companies change the price and features of their products over time, etc. All of these factors can cause pairwise comparison matrices to change over time~\cite{maystre2019pairwise,glickman1993paired,saaty2007time}. Additionally, data may arrive in streaming fashion, noisy and incomplete. For such scenarios, it is important to develop inference techniques that do not require retraining on a fresh, complete data set but rather can naturally update their estimates of the matrix/features in streaming fashion. Consequently, the goal of this paper is to recover time-varying non-transitive pairwise comparison matrices from noisy partial/incomplete observations.

{\bf Contribution.} We introduce a dynamic model to characterize skew-symmetry, low-rankness, and non-transitivity for time-varying non-transitive pairwise comparison matrices. We  formulate time-varying pairwise comparison matrix recovery as a dynamic skew-symmetric matrix recovery problem, and we propose a non-convex optimization program to recover the time-varying pairwise comparison matrix from its noisy linear observations. Moreover, in a matrix completion setting, we develop an upper bound for the recovery error in terms of the number of observed pairwise comparisons, the number of latent features, the number of items to be compared, the measurement noise variance, etc. We numerically test our developed theory with both synthetic and real-world data.

{\bf Organization.} The remainder of this paper is organized as follows. We propose a dynamic model for non-transitive pairwise comparison matrices and formulate the problem of time-varying non-transitive pairwise comparison matrix recovery in~\Cref{sec:prob}. We provide theoretical guarantees for the recovery of time-varying non-transitive pairwise comparison matrices in~\Cref{sec:err}. We numerically test our proposed theory with both synthetic and real-world data in~\Cref{sec:nume}. Finally, we conclude our work in~\Cref{sec:conc}.

{\bf Notation.} Throughout this work, we use non-boldface letters (e.g., $x, X$), boldface lowercase letters (e.g., $\x$), and boldface uppercase letters (e.g., $\X$) to denote scalars, vectors, and matrices, respectively. The $i$-th entry of a vector $\x$ is denoted as $x_i$. The $i$-th row, $j$-th column, and $(i,j)$-th entry of a matrix $\X$ are denoted as $\X_{i:}$, $\X_{:j}$, and $X_{ij}$, respectively. The superscript $^\top$ denotes the transpose of a matrix or vector. Define the set $[T]$ as $[T] \triangleq \{1,2,\cdots,T\}$. We use $\|\X\|_F$, $\|\X\|$, $\|\X\|_*$, and $\|\X\|_\infty = \max_{i,j}|X_{ij}|$ to denote the Frobenius norm, spectral norm, nuclear norm, and element-wise infinity norm, respectively. Let rank$(\cdot)$ denote the rank of a matrix. We use $C,c,C_1,c_1,C_2,c_2,\ldots$ to denote numerical constants that may vary from line to line.

\section{Problem Formulation}
\label{sec:prob}
To characterize the dynamic behavior and non-transitivity in a pairwise comparison matrix, we extend the model proposed in~\cite{yang2015modeling} to the following
\begin{align*}
\X^t = \S^t{\Q^t}^\top - \Q^t {\S^t}^\top,~\forall~t\in[T],
\end{align*}
where $T$ denotes the current time and $t$ denotes the time index. For simplicity, we assume the latent feature matrix $\Q\in \RRR^{N\times r}$ is a fixed matrix while $\S^t\in \RRR^{N\times r}$ changes over time; we leave for future work the fully general problem of including dynamic $\Q^t$. Then, the above model reduces to
\begin{align}
\X^t = \S^t\Q^\top - \Q {\S^t}^\top,~\forall~t\in[T].
\label{eq:X_form}
\end{align}
For a simple example, consider a scenario involving pairwise comparisons of shoes. Let $\s = \begin{bmatrix} s_1 & s_2 & \cdots & s_N \end{bmatrix}^\top$ with $s_i$ denoting the price for the $i$-th type of shoes (the lower the better). Similarly, let $\q = \begin{bmatrix} q_1 & q_2 & \cdots & q_N \end{bmatrix}^\top$ with $q_i$ denoting the comfort score for the $i$-th type of shoes (the higher the better). Then, one can formulate a pairwise comparison matrix as
\begin{align*}
	\X = \s \q^\top-\q\s^\top
\end{align*}
 with $X_{ij} = s_i q_j - q_i s_j>0$ indicating that shoe type $j$ is preferred over shoe type $i$. For example, if shoe type $j$ is both cheaper ($s_i > s_j$) and more comfortable ($q_j > q_i$) than shoe type $i$, it follows that $X_{ij} = s_i q_j - q_i s_j>0$. Now suppose that the prices ($\s$) of the shoes change over time while the comfort scores ($\q$) remain fixed. Then one can model the time-varying pairwise comparison matrix as
 \begin{align*}
 	\X^t = \s^t \q^\top-\q{\s^t}^\top,
 \end{align*}
which coincides with the model in~\eqref{eq:X_form} with $r=1$. A simplified model, which still agrees with~\eqref{eq:X_form}, is when $\q=\one$ and pairwise comparisons depend only on the time-varying prices of shoes.

Pairwise comparison data may be incomplete or noisy. In this work, we consider the problem of recovering a time-varying pairwise comparison matrix $\X^t\in\RRR^{N\times N}$ at time $T$ from its current and previous linear measurements
\begin{align}
\y^t = \AA^t(\X^t) + \z^t \in \RRR^M,~t=1,\cdots,T,
\label{eq:measurement}	
\end{align}
where $\AA^t:\RRR^{N\times N} \rightarrow \RRR^M$ with $[\AA^t(\X^t)]_m \triangleq \langle \A^t_m,\X^t \rangle$\footnote{\textcolor{black}{Note that $\A^t_m$ denotes the $m$-th measurement matrix used at time $t$.}} is a linear operator, and $\z^t$ with i.i.d. Gaussian entries following $\NN(0,\sigma_1^2)$ is the measurement noise. That is, we aim to recover $\X^T$ from $\{\y^t\}_{t=1}^T$.
To provide a foundation for analysis, we assume that the latent feature matrix $\S^t$ changes over time according to the following model
\begin{align}
\S^t = \S^{t-1} + \E^t,~t=2,\cdots,T,
\label{eq:X_evolve}
\end{align}
where we assume the entries of the latent innovation matrix $\E^t \in \RRR^{N\times r}$ are i.i.d. Gaussian random variables following $\NN(0,\sigma_2^2)$.

Observing that $\X^T$ is a skew-symmetric matrix with rank at most $2r$, time-varying pairwise comparison matrix recovery can be viewed as a dynamic low-rank skew-symmetric matrix recovery problem. In particular, we propose the following optimization program to recover $\X^T$ from  $\{\y^t\}_{t=1}^T$:
\begin{align}
\Xh^T  =  \arg  \min_{\X = \S\Q^\top  - \Q\S^\top}	  \sumt  \frac 1 2 w_t  \left\|\AA^t  \left(\S\Q^\top   -  \Q\S^\top\right)  -  \y^t \right\|_2^2,
\label{eq:def_opt}
\end{align}
where $\{w_t\}_{t=1}^T$ with $\sumt w_t = 1$ denote some non-negative weights, and where the optimization is with respect to the variables $\S,\Q\in \RRR^{N\times r}$. To solve this optimization problem, one can employ alternating minimization, which alternatively minimizes the cost function over one variable (e.g., $\S$) while fixing the other variable (e.g., $\Q$). A similar problem without the skew-symmetric structure is studied in~\cite{xu2016dynamic}. The global convergence analysis of alternating minimization for the static case has been established~\cite{hardt2014understanding,jain2013low,zhao2015nonconvex}. However, we leave for future work the extension of these results to the above dynamic framework.

\section{Theoretical Recovery Guarantees}
\label{sec:err}
Define the recovery error as
\begin{align*}
\|\Delta^T\|_F^2 = \|\Xh^T - \X^T\|_F^2,	
\end{align*}
where $\X^T$ is the ground truth and $\Xh^T$ is the estimator from the optimization program~\eqref{eq:def_opt}. Our goal in this section is to  upper bound  $\|\Delta^T\|_F^2$ under the matrix completion setting. In particular, we assume all of the linear operators $\{\AA^t\}_{t=1}^T$ are {\em uniform sampling ensembles} defined below.
\begin{Definition}\cite[Definition~3.5]{xu2016dynamic} \label{Def_A}
A linear operator $\AA:\RRR^{N\times N} \rightarrow \RRR^M$ with $[\AA(\X)]_m \triangleq \langle \A_m,\X \rangle$ is a {\em uniform sampling ensemble} (with replacement) if all the measurement matrices $\{\A_m\}_{m=1}^M$ are i.i.d. uniformly drawn from the set
\begin{align*}
\XX \triangleq \left\{\e_{n_1}\e_{n_2}^\top, 1\leq n_1,n_2 \leq N  \right\},	
\end{align*}
where $\e_n\in\RRR^N$ denotes the $n$-th column of an $N\times N$ identity matrix $\I_N$.
\end{Definition}

A typical assumption used in matrix completion is that the low-rank matrix to be recovered satisfies a certain incoherence property, which guarantees that it is far from a sparse matrix~\cite{xu2016dynamic,candes2009exact,jain2013low}. We present the definition of matrix incoherence below.
\begin{Definition}\cite{xu2016dynamic,candes2009exact,jain2013low}
A rank-$r$ matrix $\X\in\RRR^{N\times N}$ with an SVD $\X = \U \bSigma \V^\top$ is incoherent with parameter $\mu$ if
\begin{align*}
\|\e_n^\top \U\|_2\leq \frac{\mu\sqrt{r}}{\sqrt{N}}, \quad \|\e_n^\top \V\|_2\leq \frac{\mu\sqrt{r}}{\sqrt{N}}, \quad \forall~n\in[N].
\end{align*}	
\end{Definition}

As in~\cite{xu2016dynamic}, we also assume that $\|\X^T\|_\infty \leq a$, namely, the maximum entry of $\X^T$ in absolute value is bounded by a constant $a$. To enforce this spikiness constraint on $\X^T$, we modify the optimization program~\eqref{eq:def_opt} as
\begin{align}
\Xh^T  =  \arg  \min_{\substack{\X = \S\Q^\top  - \Q\S^\top\\ \|\X\|_\infty \leq a}}	  \sumt  \frac 1 2 w_t  \left\|\AA^t  \left(\S\Q^\top  -  \Q\S^\top\right)  -  \y^t \right\|_2^2.
\label{eq:def_opt_a}
\end{align}

We are now in position to state our main theorem on bounding the recovery error.
\begin{Theorem}\label{THM_error}
Consider a rank-$2r$ skew-symmetric matrix $\X^t\in\RRR^{N\times N}$ with form~\eqref{eq:X_form} that evolves according to model~\eqref{eq:X_evolve} with the entries of $\E^t$ following $\NN(0,\sigma_2^2)$. We further assume that each $\X^t$ is incoherent with parameter $\mu~(1\leq \mu\leq \sqrt{\frac{N}{2r}})$ and that $\|\X^T\|_\infty \leq a$.
Given are the measurements $\y^t$ as in~\eqref{eq:measurement} with all of the linear operators $\{\AA^t\}_{t=1}^T$ being independent uniform sampling ensembles. Suppose that the measurement noise vector $\z^t$ is a Gaussian random vector with entries following $\NN(0,\sigma_1^2)$. Then, the recovery error can be bounded as
\begin{align*}
\|\Delta ^T\|_F^2  &\leq  \max  \left\{  B_1  \triangleq  C_1 a^2N^2 \sqrt{\frac{\sumt w_t^2 \log(2N)}{M}}, B_2\right\},\\
  B_2  &\triangleq  C_2 \frac{rN^3\log(2N)}{M}   \left(  \sumt  w_t^2(2\sigma_2^2(T - t) + \sigma_1^2)  +  a^2  \sumt  w_t^2     \right)
\end{align*}
with probability at least $1 - c_1N^{-1} - c_2TN\text{{\em exp}}(-N)$
 if the number of measurements $M$ satisfies
 \begin{equation}
 \begin{aligned}
 M \geq CN\log(2TN^3)\log(2N) \cdot \frac{ \left( \frac{\sigma_2 \mu \sqrt{r}}{\sqrt{N}}\sqrt{\max_t w_t^2 (T-t)} + \sigma_1\sqrt{\max_t w_t^2} \right)^2}{ \sumt w_t^2 \left(  \sigma_1^2 +  2  (T-t)   \sigma_2^2 \right)}.
 \label{eq:M}	
 \end{aligned}
 \end{equation} 	
\end{Theorem}
The proof of~\cref{THM_error} is given in~\cref{proof_THM_error}. Recall that the number of measurements $M$ needed for perfect noiseless recovery should be at least on the order of $rN^{1.2}$ in classical matrix completion~\cite{candes2009exact}. Though no such factor of $r$ appears in the sample complexity bound~\eqref{eq:M}, we would like to clarify that the recovery error bound in~\cref{THM_error} does scale with $r$ when $B_2$ dominates. To then guarantee that this recovery error is smaller than $\|\X^T\|_F^2 \leq a^2N^2$, the number of samples $M$ needs to scale linearly with $rN$, as in the classical matrix completion problem.

It can be seen that the recovery bound in~\cref{THM_error} is comparable to the dynamic low-rank matrix completion results given in~\cite[Theorem 3.8]{xu2016dynamic}.
If we set $w_T=1$ and $w_t=0$ for all $t\in[T-1]$, the recovery bound in~\cref{THM_error} reduces to
\begin{equation}
\begin{aligned}
\|\Delta ^T\|_F^2  \leq  \max  \left\{   C_1 a^2N^2 \sqrt{\frac{\log(2N)}{M}}, ~  C_2 \frac{rN^3\log(2N)}{M}  \left( \sigma_1^2 + a^2    \right)     \right\},
\label{eq:wT=1}
\end{aligned}
\end{equation}
which is comparable to the {\em classical (static)} skew-symmetric matrix completion results given in~\cite[Theorem 2.9]{chen2020nonconvex}. Though the problem setup in \cite{chen2020nonconvex} is a bit different, both our bound~\eqref{eq:wT=1} and theirs are on the order of $rN^3\sigma_1^2/M$.

In a special case when $\sigma_2 = 0$, i.e., $\{\S^t\}_{t=1}^T$ are the same. Then, the bound in~\cref{THM_error} reduces to
\begin{equation}
\begin{aligned}
\|\Delta ^T\|_F^2  \leq & \max  \left\{   C_1 a^2N^2 \sqrt{\frac{\sumt w_t^2 \log(2N)}{M}}, ~   C_2 \frac{rN^3\log(2N)}{M}   \sumt w_t^2\left(\sigma_1^2 + a^2    \right)     \right\},
\label{eq:err_reduce1}
\end{aligned}
\end{equation}
which implies that the optimal weights should be $w_t = \frac 1 T$ for all $t\in[T]$. In another special case when $\sigma_2$ is large, i.e., $\S^t$ changes dramatically when compared with $\S^{t-1}$, one would expect to use only $\y^T$ to recover $\X^T$. That is, one should set $w_T=1$ and $w_t=0$ for all $t\in[T-1]$. For the general case, to find the optimal weights that minimize the recovery bound in ~\cref{THM_error}, we propose to solve the following optimization program
\begin{align*}
\{w_t^\star\}_{t=1}^T &=  \arg  \min_{\substack{ \sumt   w_t = 1\\ w_t \geq 0}}  \sumt  w_t^2(2\sigma_2^2(T - t) + \sigma_1^2)  +  a^2  \sumt   w_t^2\\
&=  \arg \min_{\substack{ \sumt w_t = 1\\ w_t \geq 0}} \sumt w_t^2\left(\frac{2\sigma_2^2}{\sigma_1^2 + a^2}(T-t)+1\right).
\end{align*}
With some fundamental calculations, one can get the analytical solution
\begin{align}
	w_t^\star = \frac{1}{\sum_{j=1}^T \frac{1}{1+\frac{2\sigma_2^2}{\sigma_1^2 + a^2}(T-j)}}\frac{1}{1+\frac{2\sigma_2^2}{\sigma_1^2 + a^2}(T-t)}, ~\forall~t\in[T],
	\label{eq:opt_w}
\end{align}
which also confirms the two special cases discussed above. In particular, we have $w_t^\star = \frac 1 T$ when $\sigma_2=0$, and $w_t^\star$ converges to $w_T^\star = 1$, $w_t^\star = 0$ for all $t\in[T-1]$ when $\sigma_2$ is sufficiently large. An alternative way to compute the optimal weights is to solve the following optimization program
 \begin{align*}
\{w_t^\star\}_{t=1}^T = \arg \min_{\substack{ \sumt w_t = 1\\ w_t \geq 0}} \sumt w_t^2\left(\frac{2\sigma_2^2}{\sigma_1^2 }(T-t)+1\right),
\end{align*}
which has an analytical solution
\begin{align}
	w_t^\star = \frac{1}{\sum_{j=1}^T \frac{1}{1+\frac{2\sigma_2^2}{\sigma_1^2 }(T-j)}}\frac{1}{1+\frac{2\sigma_2^2}{\sigma_1^2 }(T-t)}, ~\forall~t\in[T].
	\label{eq:opt_w_na}
\end{align}
In the two special cases discussed above ($\sigma_2=0$ and $\sigma_2$ sufficiently large), these weights also agree with those prescribed by~\eqref{eq:opt_w}. However, we empirically find that the weights~\eqref{eq:opt_w_na} can generally achieve a better performance, so we use these weights in our simulations.

We also note that, in the special case when $\sigma_2 = 0$, plugging in $\{w_t = 1/T\}_{t=1}^T$, the sample complexity given in~\eqref{eq:M} reduces to
 \begin{align*}
 M \geq C\frac 1 T N\log(2TN^3)\log(2N),
 \end{align*}
which implies that the number of measurements needed at each time linearly decreases as we increase the total time duration $T$. In addition, the error bound in~\eqref{eq:err_reduce1} further reduces to
 \begin{align*}
\|\Delta ^T\|_F^2  \leq  \max  &\left\{   C_1 a^2N^2 \sqrt{\frac{\log(2N)}{MT}}, ~   C_2 \frac{rN^3\log(2N)}{MT}   \left(\sigma_1^2 + a^2    \right)     \right\}.
\end{align*}
Note that the factor $r$ in the above error bound can be offset if we control $M$ to be proportional to $r$.
On the other hand, when $\sigma_2$ is sufficiently large such that the optimal weights obtained from~\eqref{eq:opt_w} or \eqref{eq:opt_w_na} are $w_T^\star = 1$, $w_t^\star = 0$ for all $t\in[T-1]$, the sample complexity bound given in~\eqref{eq:M} reduces to
 \begin{align*}
 M \geq C  N\log(2TN^3)\log(2N),
 \end{align*}
which, as expected, no longer scales with $1/T$.

\section{Numerical Simulations}
\label{sec:nume}
In this section, we conduct a series of experiments on both synthetic data and real-world data to show the performance of our proposed model and further support our theoretical analysis in the problem of time-varying pairwise comparison matrix recovery, \textcolor{black}{which is formulated as in~\eqref{eq:def_opt} and is solved with alternating minimization.}

\subsection{Synthetic data}
\begin{figure}[t]
\begin{minipage}{0.49\linewidth}
\centering
\includegraphics[width=2.8in]{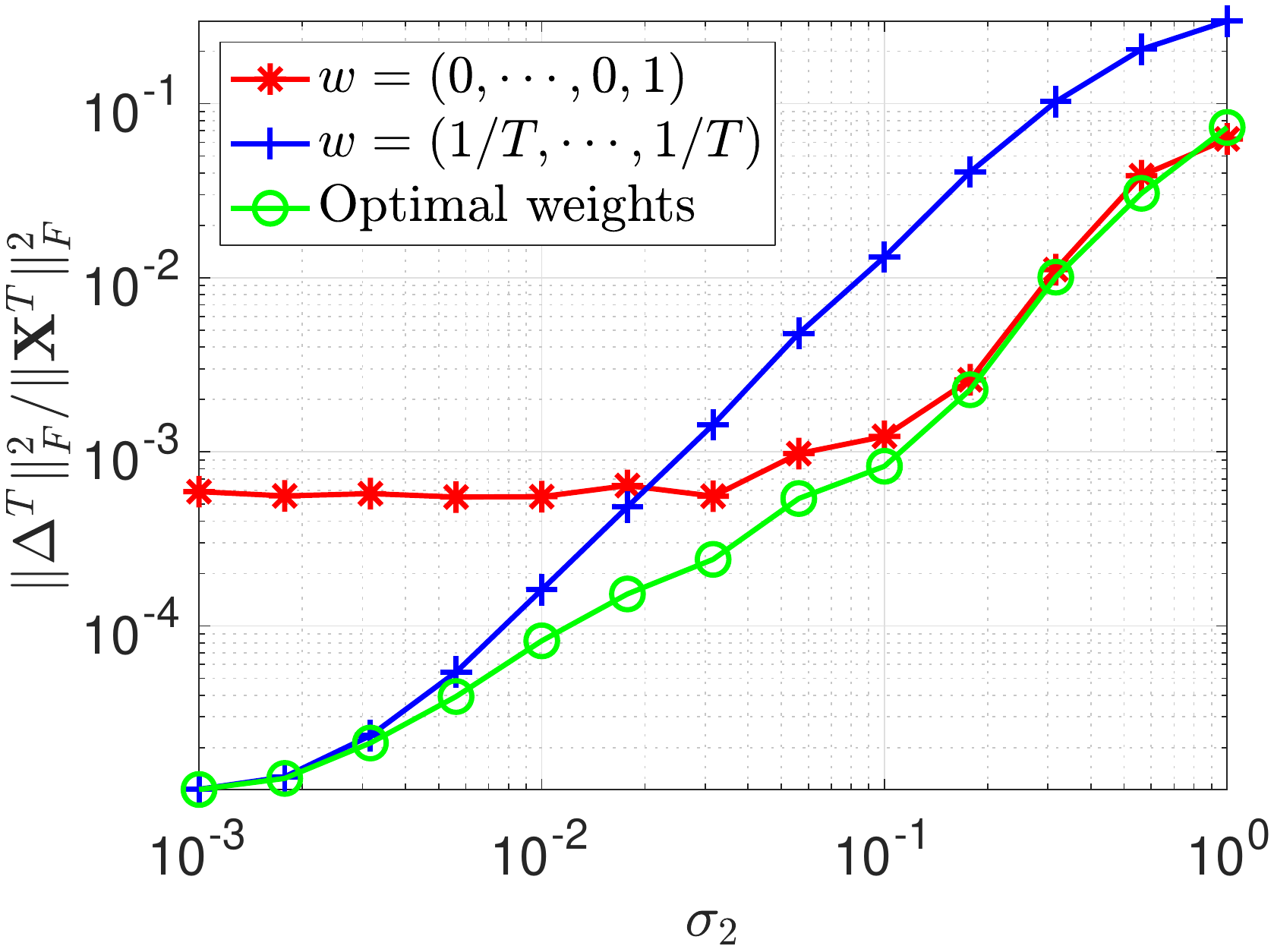}
\centerline{\small{(a)}}
\end{minipage}
\begin{minipage}{0.49\linewidth}
\centering
\includegraphics[width=2.8in]{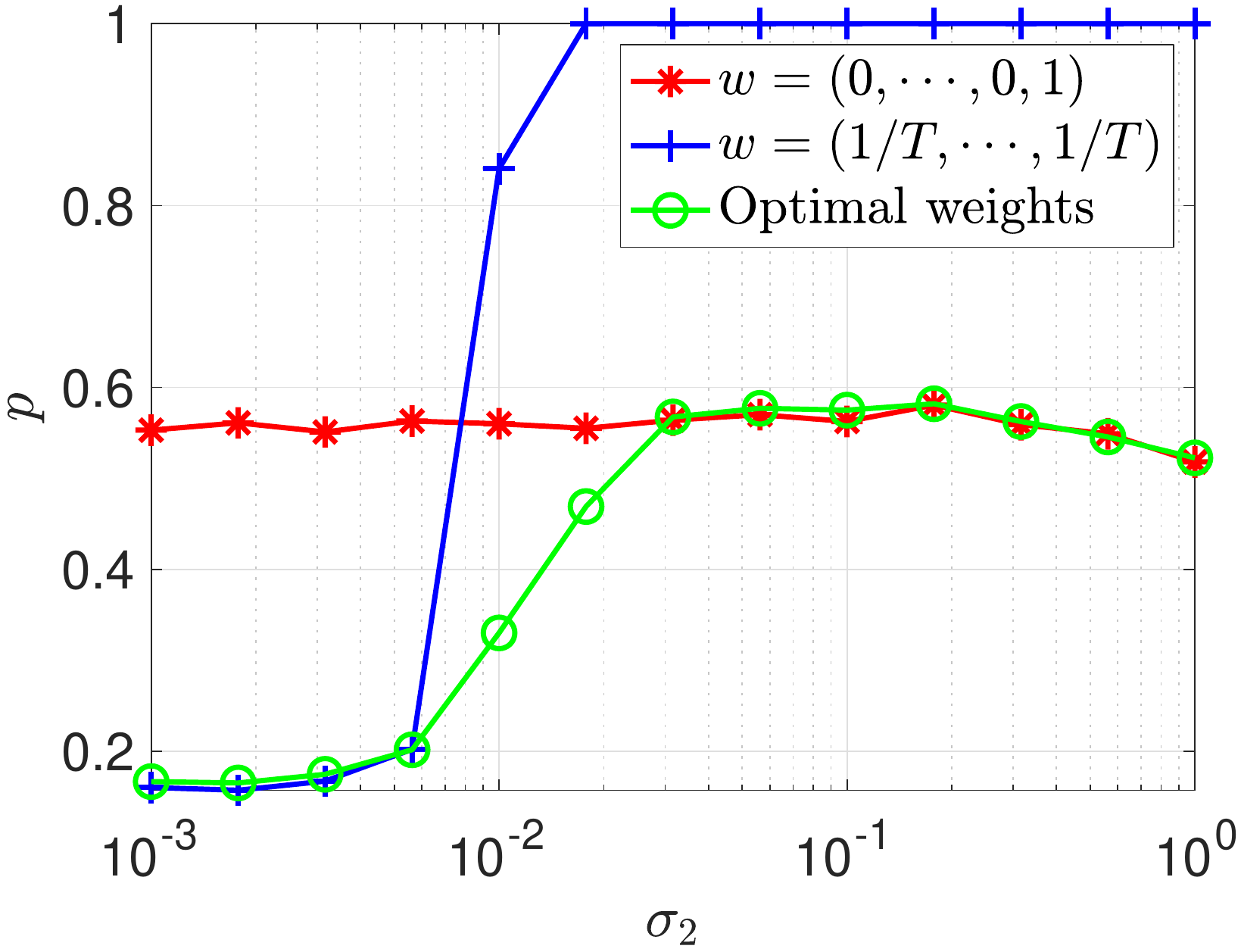}
\centerline{\small{(b)}}
\end{minipage}
\caption{Dynamic skew-symmetric matrix completion: (a) relative recovery error with respect to different $\sigma_2$; (b) sampling ratio needed for successful recovery with respect to different $\sigma_2$.}
\label{fig:test_DSSM_sig2}
\end{figure}

In this experiment, we fix $N = 20$, $r = 2$, $T = 5$, and $\sigma_1 = 0.01$. We set the sampling rate $p=0.4$, i.e., $M = pN^2 = 160$.
We generate $\Q\in\RRR^{N\times r}$ and $\S^1\in\RRR^{N\times r}$ as two random Gaussian matrices with entries following $\NN(0,1)$. $\{\S^t\}_{t=2}^T$ and $\{\X^t\}_{t=1}^T$ are generated according to the models~\eqref{eq:X_evolve} and~\eqref{eq:X_form}, respectively.
Then, we generate the linear measurements $\{\y^t\}_{t=1}^T$  in~\eqref{eq:measurement} with uniform sampling ensembles $\{\AA^t\}_{t=1}^T$ and a set of varying $\sigma_2$. The result is averaged over 100 trials. We present the relative recovery error
$$\frac{\|\Xh^T-\X^T\|_F^2}{\|\X^T\|_F^2} = \frac{\|\Delta^T\|_F^2}{\|\X^T\|_F^2}$$
in~\Cref{fig:test_DSSM_sig2}(a). Here, we test our method with three different weights: 1) $w=(0,\cdots,0,1)$: $w_T=1$, $w_t=0$ for all $t\in[T-1]$, that is, recovering $\X^T$ with only $\y^T$ and ignoring $\{\y^t\}_{t=1}^{T-1}$; 2) $w = (1/T,\cdots,1/T)$: $w_t=1/T$ for all $t\in[T]$, that is, recovering $\X^T$ with equal weights on $\{\y^t\}_{t=1}^T$; and 3) optimal weights computed via~\eqref{eq:opt_w_na}. It can be seen that when $\sigma_2$ is small, the relative recovery error obtained from using optimal weights converges to the one using equal weights $w = (1/T,\cdots,1/T)$. On the other hand, when $\sigma_2$ is large, the relative recovery error obtained from using optimal weights converges to the one using weights $w=(0,\cdots,0,1)$. This coincides with the analysis presented in~\Cref{sec:err}.

Next, we repeat the above experiment by varying the sampling ratio $p = M/N^2$ and empirically show the relationship between $\sigma_2$ and the sample complexity needed for successful recovery. Here, we define the recovery to be a success if $$\frac{\|\Delta^T\|_F^2}{\|\X^T\|_F^2} \leq 10^{-4}.$$
The result shown in~\Cref{fig:test_DSSM_sig2}(b) is averaged over 100 trials. As in~\Cref{fig:test_DSSM_sig2}(a), the sampling ratio needed for successful recovery using optimal weights converges to the one using equal weights $w = (1/T,\cdots,1/T)$ when $\sigma_2$ is small, and converges to the one using weights $w=(0,\cdots,0,1)$ when $\sigma_2$ is large. Moreover, \Cref{fig:test_DSSM_sig2}(b) also indicates that the proposed dynamic skew-symmetric matrix completion can reduce the sampling ratio needed for successful recovery when compared with the {\em static} skew-symmetric matrix completion, i.e., the case with using weights $w=(0,\cdots,0,1)$, especially when $\sigma_2$ is small.

\begin{figure}[t]
\begin{minipage}{0.49\linewidth}
\centering
\includegraphics[width=2.8in]{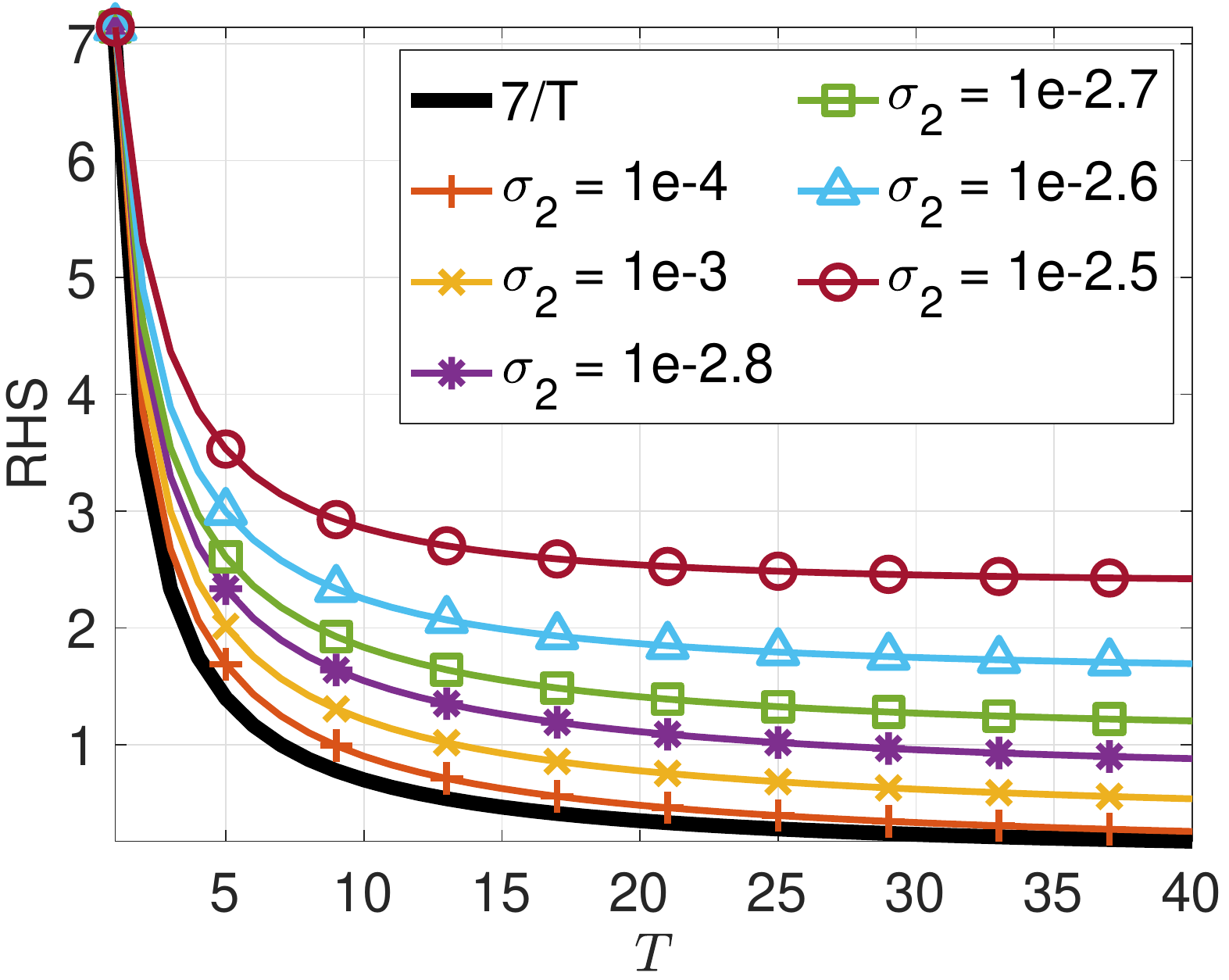}
\centerline{\small{(a)}}
\end{minipage}
\begin{minipage}{0.49\linewidth}
\centering
\includegraphics[width=2.8in]{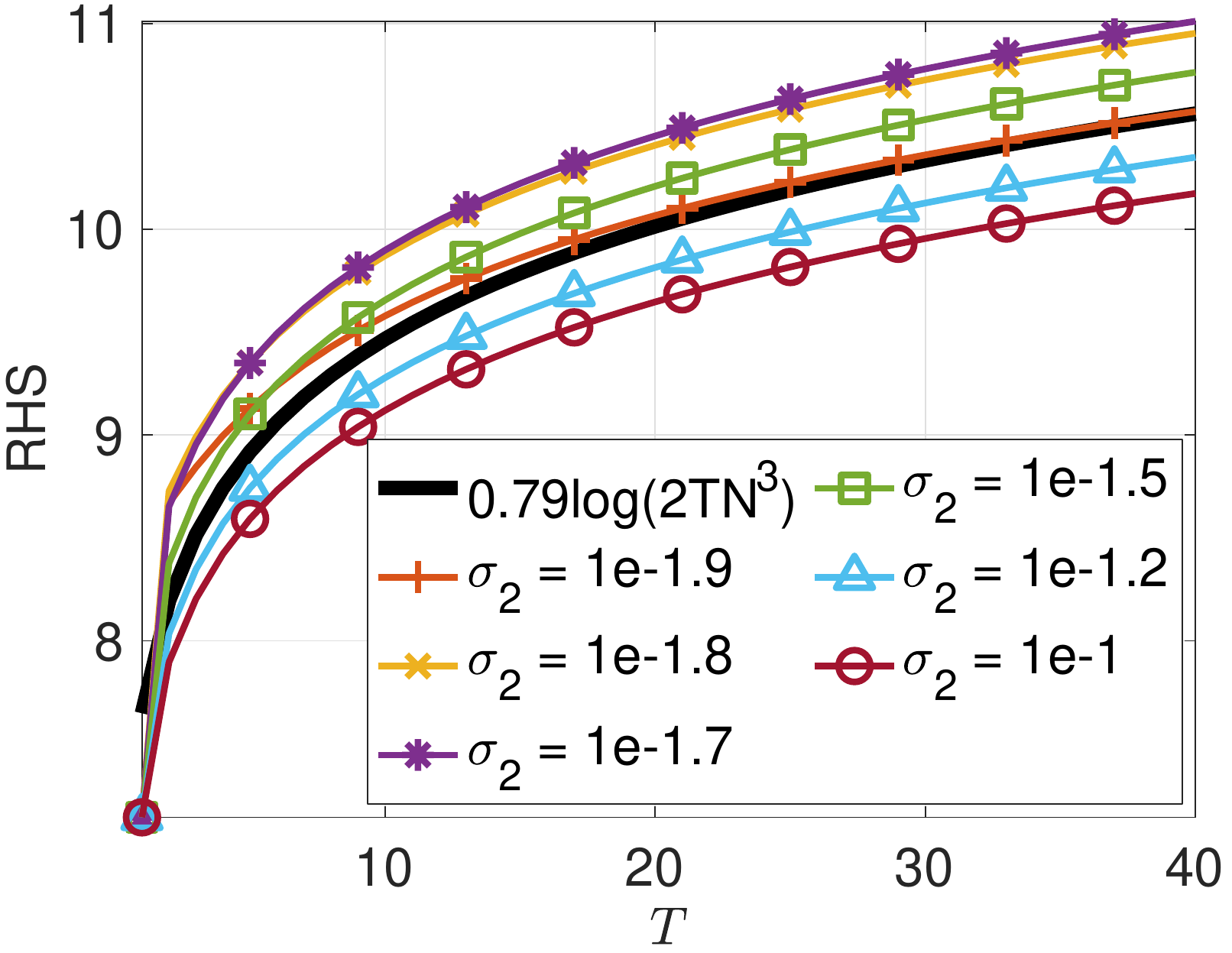}
\centerline{\small{(b)}}
\end{minipage}
\caption{Sample complexity bounds evaluated at different time duration $T$ and perturbation noise level $\sigma_2$: (a) small $\sigma_2$ and (b) large $\sigma_2$. }
\label{fig:test_DSSM_M_T}
\end{figure}

Next, we set $N=20$, $r=2$, $\mu=1$, and $\sigma_1=0.01$ and illustrate how the sample complexity bound given in the right hand side of~\eqref{eq:M} behaves with different time durations $T$ and perturbation noise levels $\sigma_2$. We define
\begin{align*}
\text{RHS}= N\log(2TN^3)\log(2N) \cdot \frac{ \left( \frac{\sigma_2 \mu \sqrt{r}}{\sqrt{N}}\sqrt{\max_t w_t^2 (T-t)} + \sigma_1\sqrt{\max_t w_t^2} \right)^2}{ \sumt w_t^2 \left(  \sigma_1^2 +  2  (T-t)   \sigma_2^2 \right)}.
\end{align*}
Note that the solid black line in~\Cref{fig:test_DSSM_M_T} denotes a baseline of either (a) $7/T$ or (b)~$0.79\log(2TN^3)$. It can be seen that when $\sigma_2$ is small enough, e.g., $\sigma_2=10^{-4}$, the sample complexity bound roughly scales as $1/T$. However, when $\sigma_2$ is sufficiently large, e.g., $\sigma_2 = 10^{-1.5}$, the sample complexity bound no longer scales with $1/T$, which coincides with the analysis provided in~\Cref{sec:err}.

\begin{figure}[t]
\begin{minipage}{0.49\linewidth}
\centering
\includegraphics[width=2.8in]{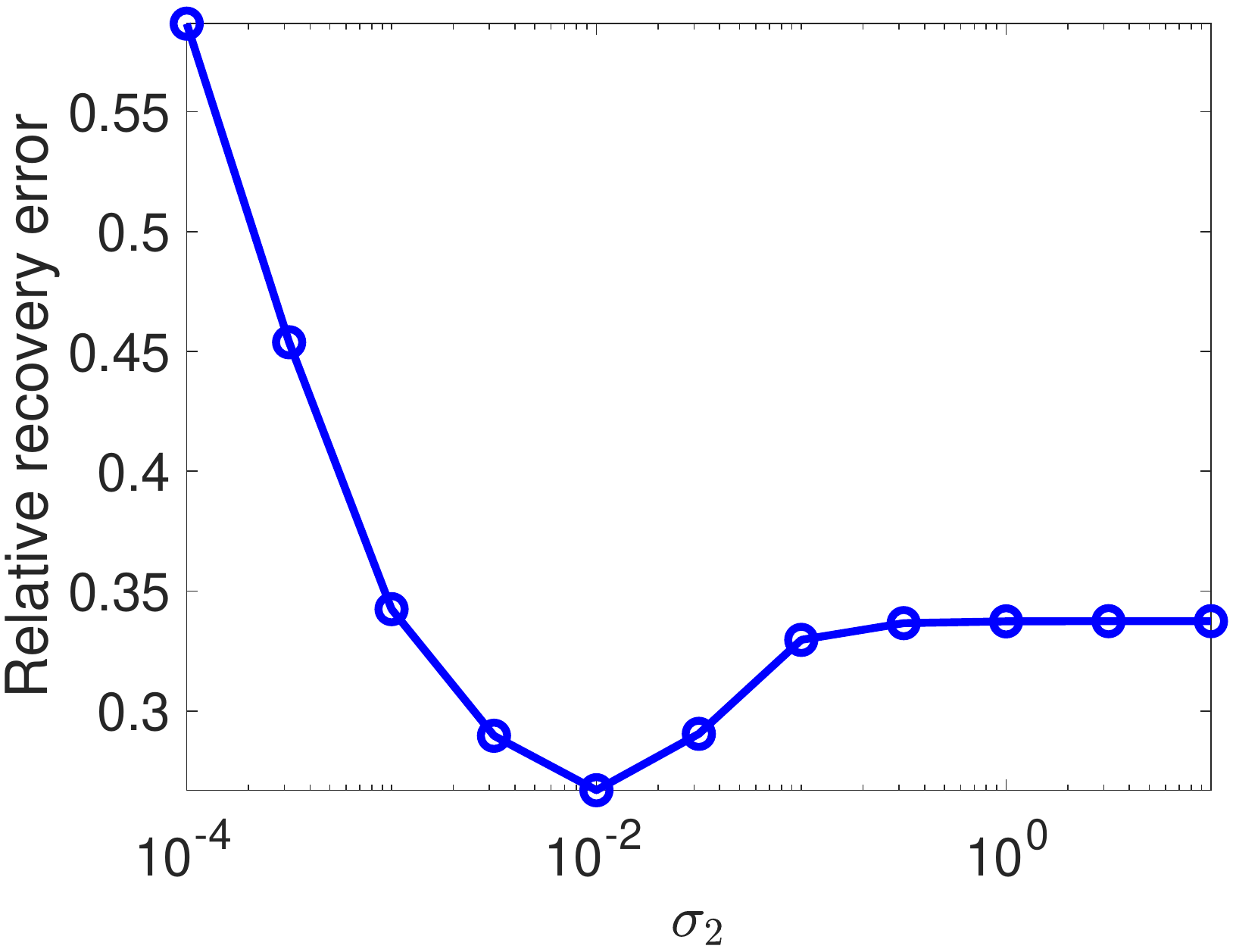}
\centerline{\small{(a)}}
\end{minipage}
\begin{minipage}{0.49\linewidth}
\centering
\includegraphics[width=2.8in]{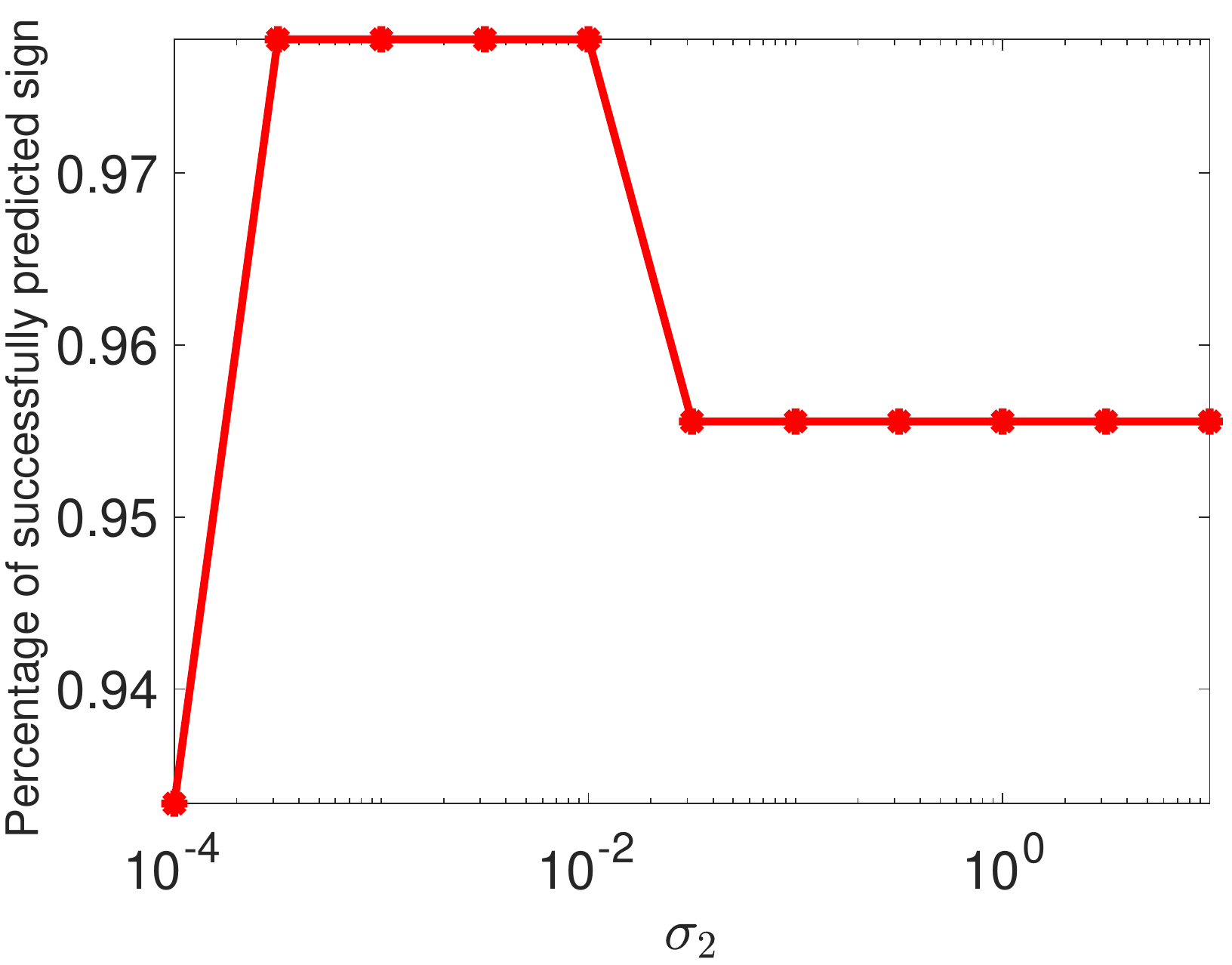}
\centerline{\small{(b)}}
\end{minipage}
\caption{Synthetic game on guessing the price of a prize: (a) relative recovery error with respect to different $\sigma_2$; (b) percentage of successfully predicted sign with respect to different $\sigma_2$.}
\label{fig:test_syntheticPW}
\end{figure}

\textcolor{black}{
Finally, we simulate a synthetic game to create a more realistic pairwise comparison matrix. To be more precise, we suppose that there are $N=10$ players. Player $i$ and player $j$ compete to guess the price $P^\star$ of a prize. The winner is the player who comes closest to the true price without going over.\footnote{These rules match those of a bidding contest on the American game show {\em The Price Is Right}.} For the sake of this game, each true price $P^\star$ is generated randomly from the uniform distribution on the interval $[0,1]$, and the players' guesses $P_i$ and $P_j$ are also generated randomly. In particular, we model each player's guess using a Beta$(a,b)$ random variable with $a$ and $b$ chosen uniformly random between 0.5 and 5. To simulate a dynamic scenario, we let each player's parameters $a$ and $b$ change slowly over time. Specifically, we add a Gaussian random number following $\NN(0,10^{-4})$ to $a$ and $b$ on the current day to get the value of $a$ and $b$ for the next day. We simulate 100 such games between each pair of players every day and repeat this over $T=10000$ days to estimate the pairwise comparison matrices $\{\X^t\}_{t=1}^T$. On day $t$, for each game between player $i$ and player $j$, we add the true price to the $(i,j)$-th entry of $\X^t$ if player $i$ wins; otherwise, we subtract the true price from the $(i,j)$-th entry of $\X^t$. We then average the entries of $\X^t$ by dividing the number of games per day (100) and fill in the bottom triangle of $\X^t$ by letting $\X^t = \X^t - {\X^t}^\top$, so that $\X^t$ is a skew-symmetric matrix. We view these estimated pairwise comparison matrices $\{\X^t\}_{t=1}^T$ over the $T=10000$ days as noisy observations of a ``ground truth'' pairwise comparison matrix $\Xs$, which is computed using $100\times 10000 = 10^6$ games between each pair of players, all on day $T$. We note that $\Xs$ is not rank $2$ and thus does not obey the transitive model~\eqref{eq:rank2trans}. Given $\{\X^t\}_{t=1}^T$, our goal is to recover an estimate of the true pairwise comparison matrix $\Xs$. In particular, we solve~\eqref{eq:def_opt} by alternating minimization and setting $r=1$. We denote the recovered $\Xs$ as $\Xh$ and present the relative recovery error $\|\Xh-\Xs\|_F/\|\Xs\|_F$ and percentage of successfully predicted signs of $\Xs$ in~\Cref{fig:test_syntheticPW}. Here, successfully predicting the sign of the $(i,j)$-th entry of $\Xs$ implies a successful prediction of the superior player. It can be seen that with the proposed model, one can achieve a low relative recovery error and high percentage of successfully predicted signs when $\sigma_2$ is appropriately chosen. (Here, $\sigma_2$ affects the algorithm via the choice of weights; it is not used in generating the data.) Importantly, $\sigma_2$ should not be chosen to small as this would cause the estimate to be corrupted by irrelevant, long past data. Nor should $\sigma_2$ be chosen too large as this would cause the estimate to ignore relevant, somewhat recent data.
}

\subsection{Real-world data}

\subsubsection{Cell counting}

\begin{figure}[t]
\begin{minipage}{0.24\linewidth}
\centering
\includegraphics[width=1.35in]{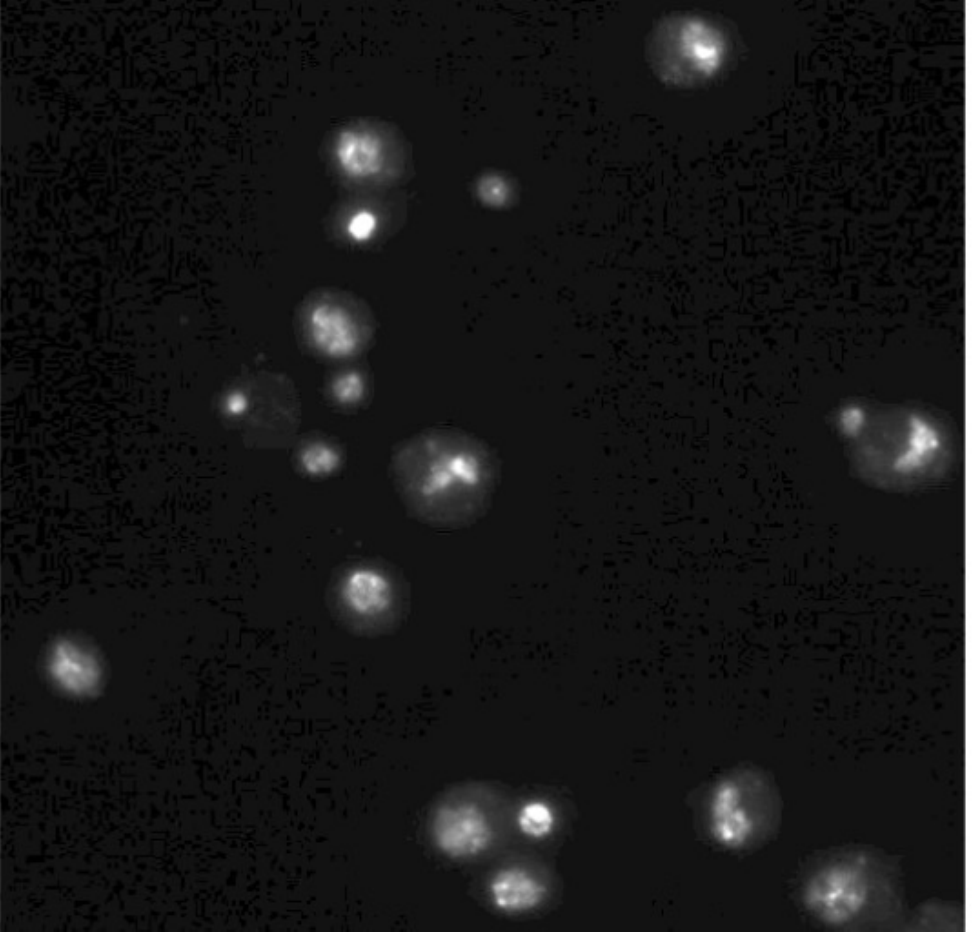}
\centerline{\small{(a)}}
\end{minipage}
\hfill
\begin{minipage}{0.24\linewidth}
\centering
\includegraphics[width=1.35in]{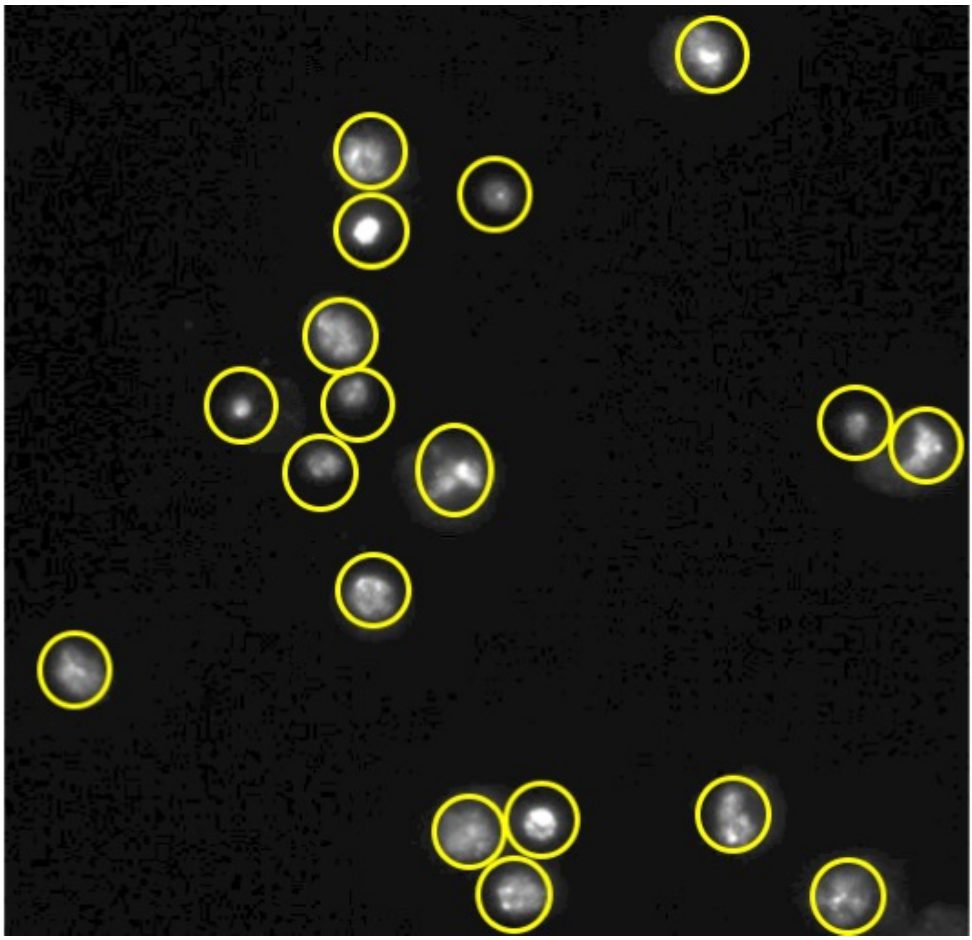}
\centerline{\small{(b)}}
\end{minipage}
\hfill
\begin{minipage}{0.24\linewidth}
\centering
\includegraphics[width=1.35in]{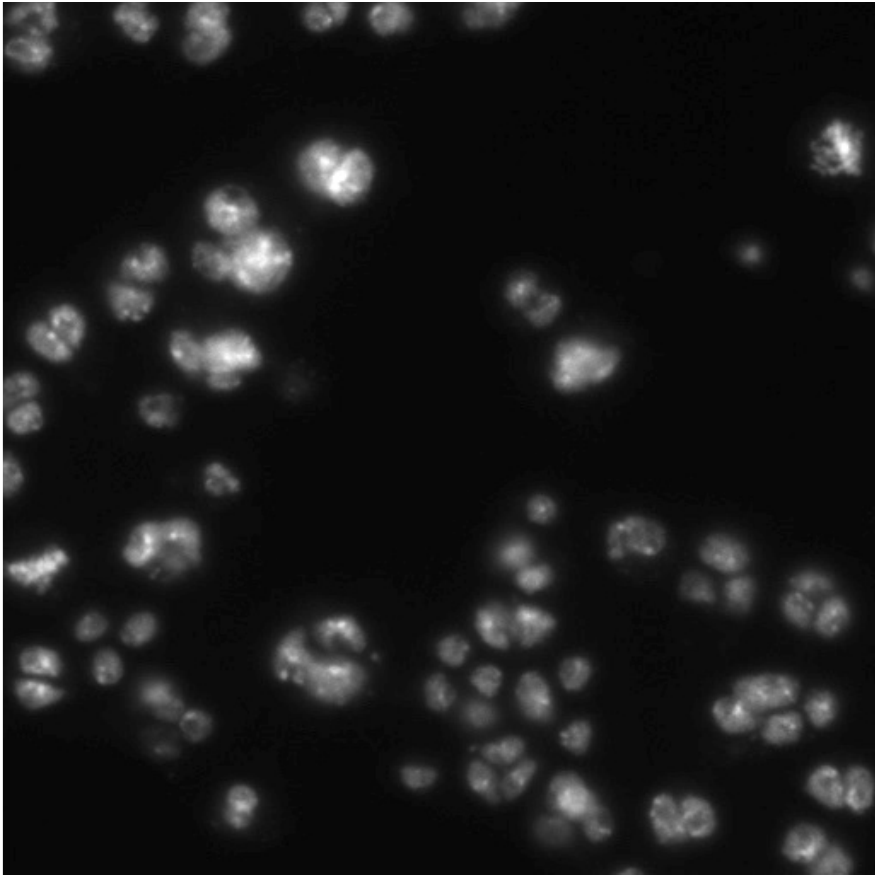}
\centerline{\small{(c)}}
\end{minipage}
~~\hfill
\begin{minipage}{0.24\linewidth}
\centering
\includegraphics[width=1.35in]{count_cells4}
\centerline{\small{(d)}}
\end{minipage}
\caption{Sample images with multiple {\em drosophila melanogaster} cells used in the cell counting problem~\cite{shah2017simple,carpenter2006cellprofiler}.}
\label{fig:cell_counting}
\end{figure}

Our next experiment is inspired by the cell counting problem~\cite{carpenter2006cellprofiler,shah2017simple}. A sample image that contains  multiple {\em drosophila melanogaster} cells is given in~\Cref{fig:cell_counting}(a) with each contiguous blob denoting a cell. To illustrate, we mark the cells with yellow circles in~\Cref{fig:cell_counting}(b). Here, we test on $N=22$ images with {\em drosophila melanogaster} cells shown in each image. We present another two sample images in~\Cref{fig:cell_counting}(c, d).\footnote{The plots in~\Cref{fig:cell_counting} are credit to a recent work~\cite{shah2017simple}. The original images and their annotation are from an earlier work~\cite{carpenter2006cellprofiler}.} From the data provided by~\cite{shah2017simple}, we can extract the true number of cells in each image. We use $\s^1\in\RRR^N$ to denote a vector that contains the number of cells in each of these 22 images. As in many rank aggregation problems~\cite{david1987ranking,hochbaum2010separation,jiang2011statistical,gleich2011rank,borda1784memoire,kemeny1959mathematics,freund2003efficient}, we can formulate a pairwise comparison matrix $\X^1\in\RRR^{N \times N}$ with $X^1_{ij} = s^1_i - s^1_j$ that encodes all possible comparisons between the numbers of cells in pairs of these 22 images. We can rewrite $\X^1$ as
$$\X^1 = \s^1 \one^\top - \one {\s^1}^\top,$$
where $\one \in \RRR^N$ is a vector containing all ones. It can be seen that the resulting pairwise comparison matrix $\X^1$ exhibits two types of structure: (1) it is low rank (actually, rank $2$), and (2)  it is skew-symmetric, satisfying the linear constraint $\X^1 = -{\X^1}^T$. Therefore, $\X^1$ obeys~\eqref{eq:rank2trans} and also fits our data model introduced in~\eqref{eq:X_form} with $t=1$ and $r=1$.

\begin{figure}[t]
\begin{minipage}{0.49\linewidth}
\centering
\includegraphics[width=2.8in]{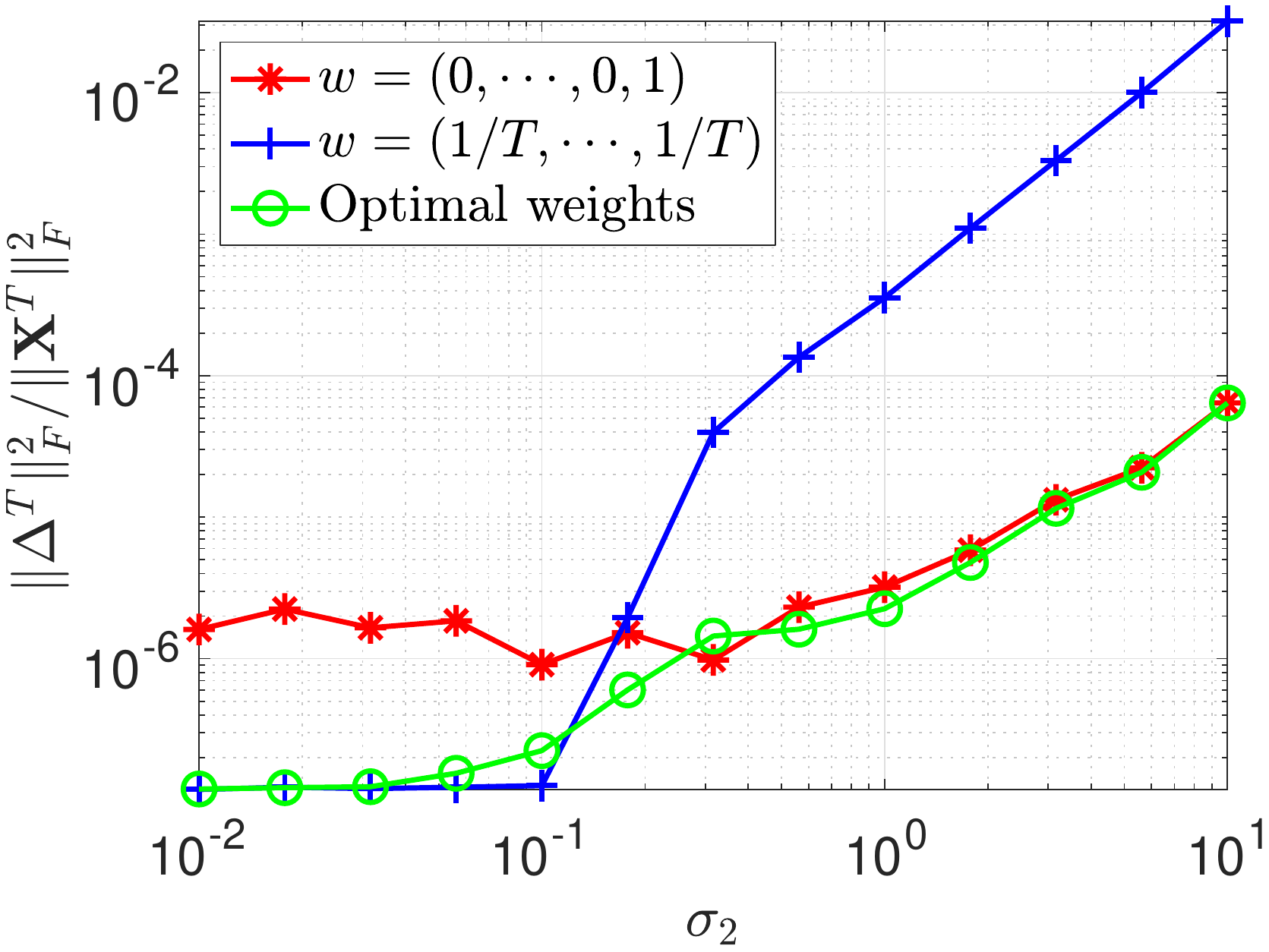}
\centerline{\small{(a)}}
\end{minipage}
\begin{minipage}{0.49\linewidth}
\centering
\includegraphics[width=2.8in]{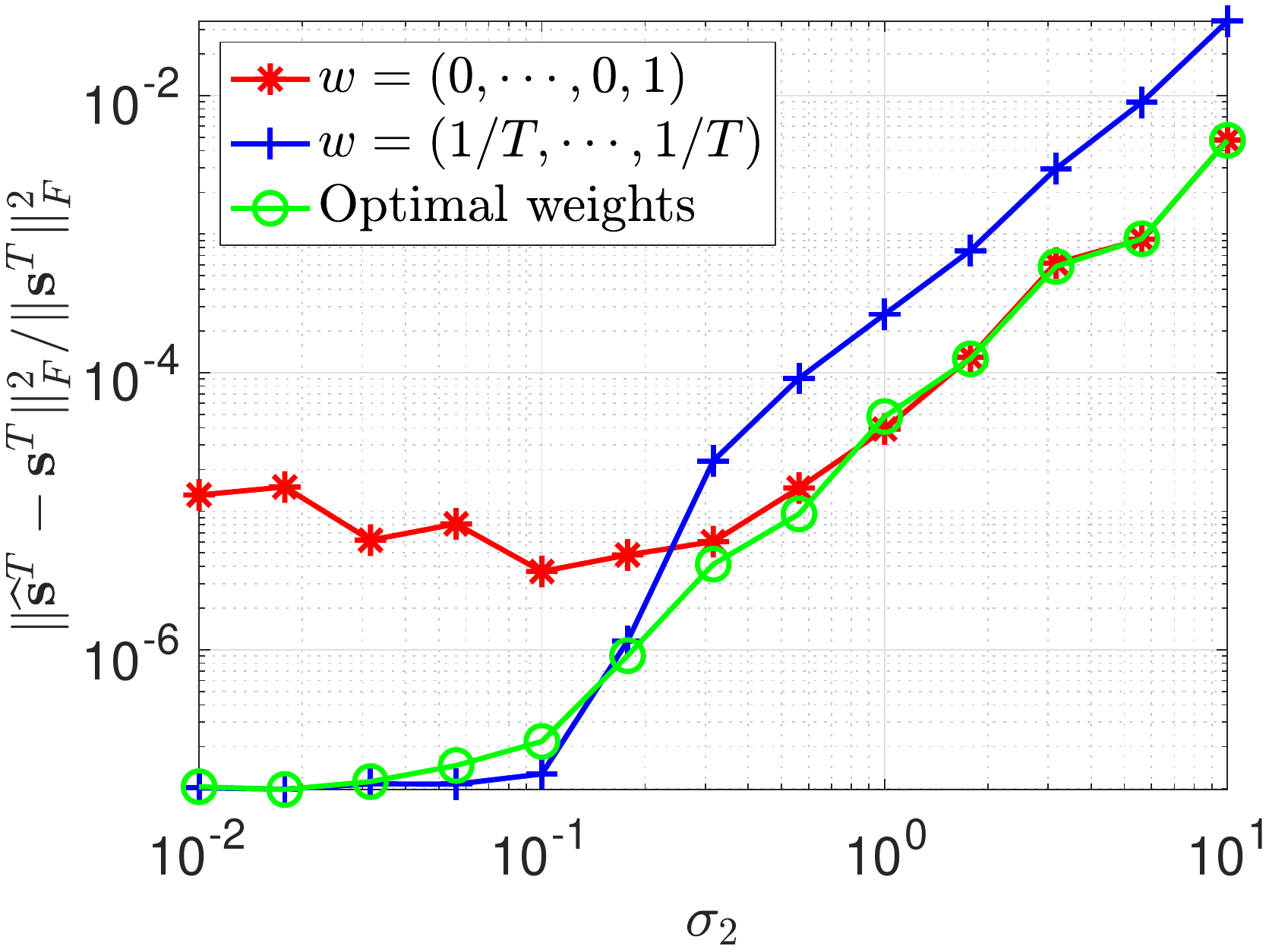}
\centerline{\small{(b)}}
\end{minipage}
\caption{Dynamic skew-symmetric matrix completion in the problem of cell counting: (a) relative matrix recovery error and (b) relative latent factor recovery error with respect to different $\sigma_2$.}
\label{fig:test_DSSM_cell}
\end{figure}

Suppose now that the number of cells in each image may change over time, which can cause the pairwise comparison matrix to change over time. To simulate this dynamic scenario, we create a series of vectors $\{\s^t\}_{t=2}^T$ according to the model introduced in~\eqref{eq:X_evolve} with $T=5$ and a set of varying $\sigma_2$. Note that we round the entries of $\E^t$ in~\eqref{eq:X_evolve} to the nearest integer before we add it to $\s^{t-1}$.  We then construct the pairwise comparison matrices $\{\X^t\}_{t=2}^T$ and generate noisy partial observations according to the linear measurement model~\eqref{eq:measurement} with $\sigma_1 = 0.1$, $p=0.2$, and $\{\AA^t\}_{t=1}^T$ being uniform sampling ensembles. Given the noisy partial observations, our goal here is to recover the latent feature vector $\s^T$, i.e., the vector that contains the number of cells in each image at time $T$. In practice, this can be used to identify the images having the fewest cells, which can be useful in real applications. For example, a low count of red blood cells in the images of human cells may indicate anemia. We present the relative recovery error for both the pairwise comparison matrix $\X^T$ and the latent feature vector $\s^T$ in~\Cref{fig:test_DSSM_cell}. The results are averaged over 100 trials. Again, we can see that the relative recovery error obtained with optimal weights converges to the one using equal weights $w = (1/T,\cdots,1/T)$ when $\sigma_2$ is small, and converges to the one using weights $w=(0,\cdots,0,1)$ when $\sigma_2$ is large.

\subsubsection{Housing price}

\begin{figure}[t]
\begin{minipage}{0.49\linewidth}
\centering
\includegraphics[width=2.8in]{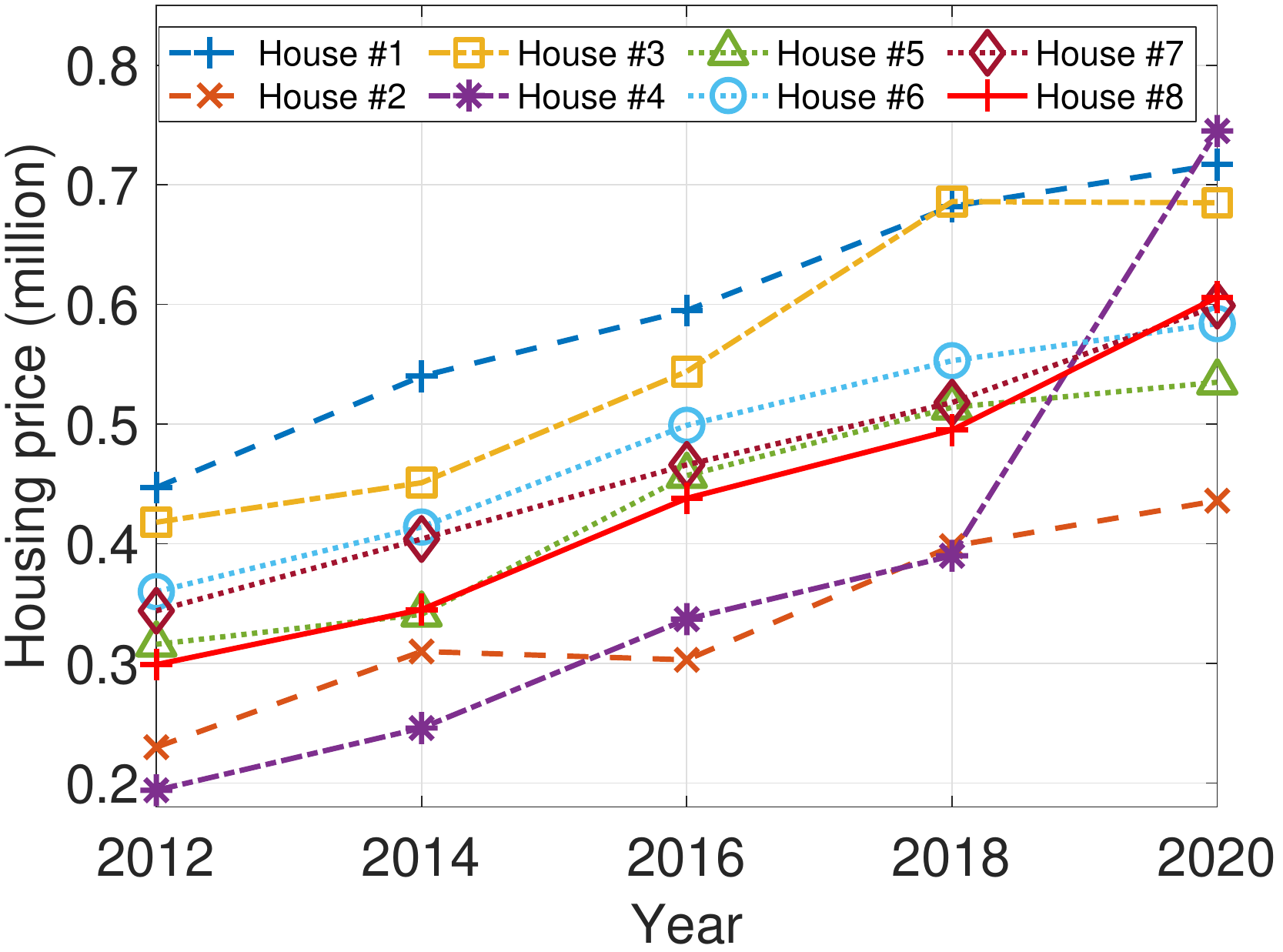}
\centerline{\small{(a)}}
\end{minipage}
\begin{minipage}{0.49\linewidth}
\centering
\includegraphics[width=2.8in]{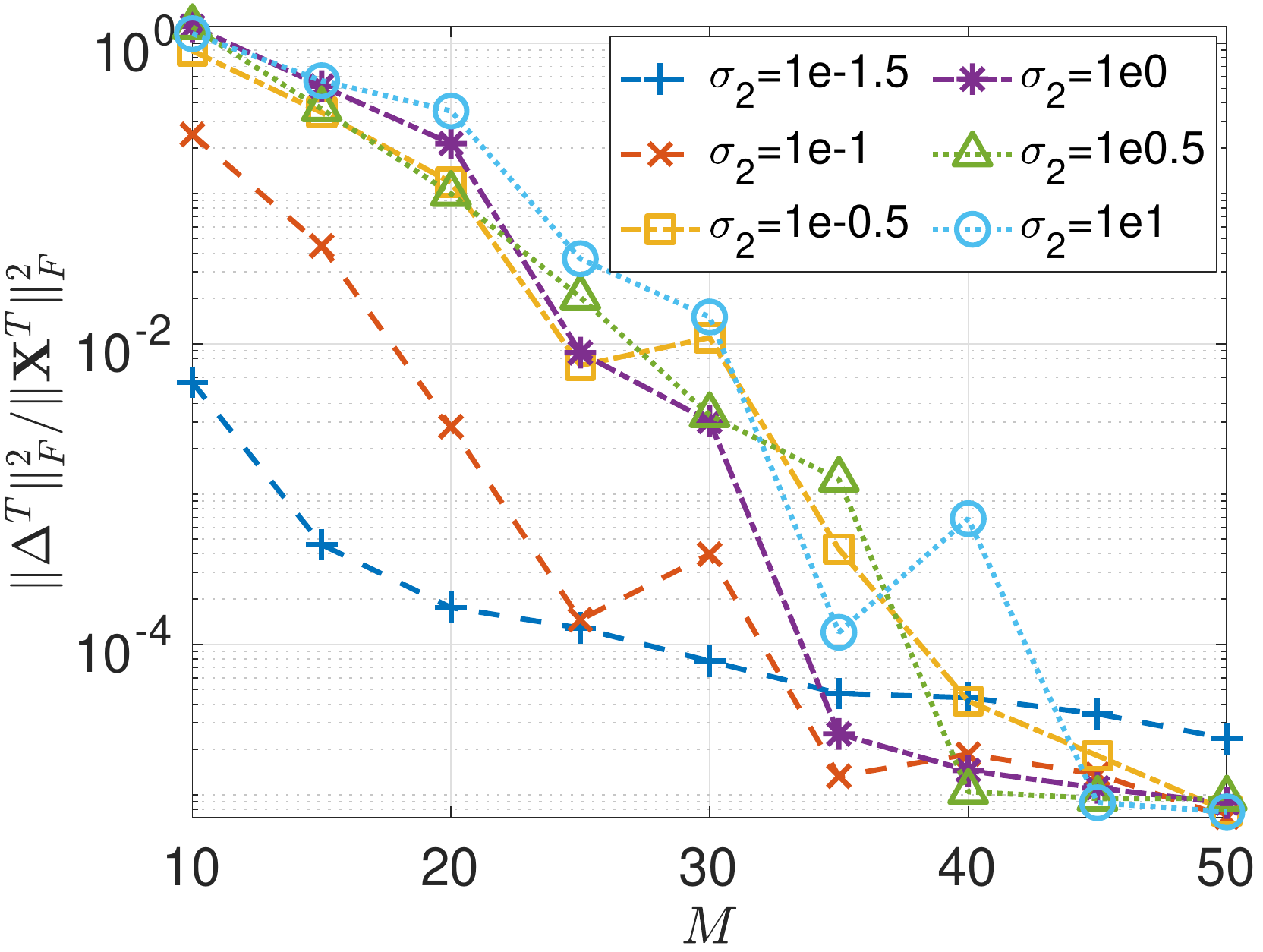}
\centerline{\small{(b)}}
\end{minipage}
\caption{Dynamic skew-symmetric matrix completion in the problem of housing price: (a)~housing price and (b) relative matrix recovery error.}
\label{fig:test_DSSM_house}
\end{figure}

For the last experiment, we test the proposed method in a scenario involving pairwise comparisons of homes. We select $N=8$ houses from Zillow and present part of the data used in this experiment in Table~\ref{table_house}. We record the prices of these houses in January every two years from 2012 to 2020 ($T=5$); see~\Cref{fig:test_DSSM_house}(a). Then, we formulate the latent feature matrix $\Q\in\RRR^{N\times r}$ ($r=2$) with its first and second column denoting the number of rooms and size of each house, respectively. Similarly, we formulate the other latent feature matrix $\S\in\RRR^{N\times r}$ with its first and second column denoting the price and age in years of each house, respectively. It can be seen that $\Q$ is fixed while $\S$ changes over time. We use $\S^t \in\RRR^{N\times r}$ to denote $\S$ at time $t$ for any $t\in[T]$. For example, $t=1$ corresponds to year 2012. Then, we formulate a set of pairwise comparison matrices $\{\X^t\}_{t=1}^T$ according to the model introduced  in~\eqref{eq:X_form} and generate noisy partial observations $\{\y_t\}_{t=1}^T$ according to the linear measurement model~\eqref{eq:measurement} with $\sigma_1 = 0.01$, and $\{\AA^t\}_{t=1}^T$ being uniform sampling ensembles. Given the noisy partial observations $\{\y_t\}_{t=1}^T$, our goal here is to recover the pairwise comparison matrix $\X^T$. However, because the latent feature matrix $\S$ used in the experiment may not exactly follow the transition model~\eqref{eq:X_evolve}, we do not have any information regarding $\sigma_2$ to use in computing the optimal weights. To overcome this, we test the proposed method with optimal weights obtained from a collection of possible assumed $\sigma_2$ values. The relative recovery errors for the pairwise comparison matrix $\X^T$ with different numbers of observations are presented in~\Cref{fig:test_DSSM_house}(b). Each result is again averaged over 100 trials. It can be seen that the proposed method can recover the pairwise comparison matrix $\X^T$ very well when provided with enough observations. In addition, the assumed value of $\sigma_2$ does have some effect on the recovery error. Moreover, when $M$ is small, smaller assumed $\sigma_2$ gives optimal weights close to equal weights, which allows the algorithm to make full use of all the available data. Thus we observe that the smallest $\sigma_2 = 10^{-1.5}$ outperforms the other $\sigma_2$ values when $M$ is small.

\begin{table}[t]
\begin{center}
\begin{tabular}{|c|c|c|c|}
\hline
& Year built & Size (sqft)  &  $\#$ of rooms \\
\hline
      House $\#1$&2003&5546&4\\
\hline
      House $\#2$&1940&3849&4\\
\hline
House $\#3$&1957&3792&6\\
\hline
      House $\#4$&1950&2484&5\\
\hline
House $\#5$&1973&2673&5\\
\hline
      House $\#6$&1999&3057&4\\
\hline
House $\#7$&1993&3744&5\\
\hline
      House $\#8$&1962&2942&3\\
\hline
\end{tabular}
\end{center}
\caption{Information for 8 houses selected from Zillow.} \label{table_house}
\end{table}

\section{Conclusions}
\label{sec:conc}

In this work, we have proposed a structured model to characterize the non-transitivity in time-varying low-rank pairwise comparisons. Given linear noisy observations under this model, we have reformulated the problem of time-varying pairwise comparison matrix recovery as one of dynamic skew-symmetric matrix recovery. We have derived an upper bound on the recovery error in the matrix completion setting and supported this analysis with a series of numerical experiments on both synthetic and real-world data.

A number of open questions could motivate future work. First, our model and analysis account only for dynamics in $\S^t$; a fully general treatment would also consider dynamics in $\Q^t$. It also remains to establish convergence analysis for the alternating minimization algorithm. In addition, we have assumed that the number of measurements is the same at each time step; it would be interesting to consider the case where the number of measurements is time-varying. Finally, in practice, the dynamic model~\eqref{eq:X_evolve} may not hold exactly or, if it does, the parameter $\sigma_2$ may be unknown as in the housing price experiment. While our method nevertheless performs well in recovering the pairwise comparison matrix, it would be valuable to support this with theoretical analysis.

\bibliographystyle{ieeetr}
\bibliography{references}

\appendix

\section{Proof of~\cref{THM_error}}
\label{proof_THM_error}

In this section, we present the proof of~\cref{THM_error}, which is inspired by the proof of \cite[Theorem~3.8]{xu2016dynamic}.\footnote{\textcolor{black}{Though our proof is inspired from \cite{xu2016dynamic}, the extra skew-symmetric structure in this work makes the extension non-trivial, especially for bounding the first term in~\eqref{eq:proof_mid1}.}}
It follows from \cite[Proposition~3.1]{xu2016dynamic} that the estimator $\Xh^T$ obtained from~\eqref{eq:def_opt_a} satisfies
\begin{align}
\sumt w_t \|\AA^t(\Delta^T)\|_2^2 \leq 4\sqrt{r}\left\| \sumt w_t {\AA^t}^* (\h^t - \z^t)  \right\|	 \|\Delta^T\|_F,
\label{eq:proof_prop}
\end{align}
where $\h^t = \AA^t(\X^T-\X^t)$ and ${\AA^t}^*:\RRR^M \rightarrow \RRR^{N\times N}$ is the adjoint operator of $\AA^t$ defined as ${\AA^t}^*(\z^t) = \summ z^t_m \A_m^t$. Note that this is a deterministic bound that holds for any set of $\{\AA^t\}_{t=1}^T$. Next, we will lower bound the LHS of~\eqref{eq:proof_prop} and upper bound the RHS of~\eqref{eq:proof_prop} for the uniform sampling ensembles $\{\AA^t\}_{t=1}^T$.

Define a set
\begin{align*}
\EE(r) \triangleq &\left\{\X\in\RRR^{N\times N}: \text{rank}(\X)\leq r, \|\X\|_\infty = 1,~\|\X\|_F^2 \geq 8N^2 \sqrt{\frac{\sumt w_t^2 \log(2N)}{\log{(6/5)}M}}   \right\}.	
\end{align*}
Let $p = \frac{M}{N^2}$ denote the sampling rate.
Suppose $\{\AA^t\}_{t=1}^T$ are fixed uniform sampling ensembles. It follows from~\cite[Lemma E.1]{xu2016dynamic} that
\begin{align}
\sumt w_t \|\AA^t(\X)\|_2^2 \geq \frac{p}{2} \|\X\|_F^2 - \frac{44r}{p}(\EEE( \|\bSigma_R\|  ))^2
\label{eq:LHS_low}	
\end{align}
holds for all $\X\in\EE(r)$ with probability at least $1-N^{-1}$. Here, $\bSigma_R \in \RRR^{N\times N}$ is a random matrix defined as
\begin{align*}
\bSigma_R \triangleq \sumt \summ w_t \gamma_m^t \A_m^t	
\end{align*}
with $\gamma_m^t$ being Rademacher variables.
Note that $\|\Delta^T\|_\infty \leq \|\Xh^T\|_\infty + \|\X^T\|_\infty \leq 2a$,  rank$(\frac{\Delta^T}{2a}) \leq 4r$, and $\|\frac{\Delta^T}{2a}\|_\infty \leq 1$. To proceed, we consider the following two cases.

\vspace{0.3cm}
\noindent
{\bf Case I:} $\frac{\Delta^T}{2a} \notin \EE(4r)$.
 According to the definition of $\EE(4r)$, we immediately get
\begin{align*}
\|\Delta^T\|_F^2 \leq C_1 a^2N^2 \sqrt{\frac{\sumt w_t^2 \log(2N)}{M}},
\end{align*}
where $C_1 = \frac{32}{\sqrt{\log{(6/5)}}}$ is a numerical constant.

\vspace{0.3cm}
\noindent
{\bf Case II:} $\frac{\Delta^T}{2a} \in \EE(4r)$. It follows from~\eqref{eq:LHS_low} that
\begin{align*}
\sumt w_t \|\AA^t(\Delta^T)\|_2^2 \geq \frac{p}{2} \|\Delta^T\|_F^2 - \frac{704r}{p}(\EEE( \|\bSigma_R\|  ))^2 a^2.
\end{align*}
Together with~\eqref{eq:proof_prop}, we have
\begin{align*}
\frac{p}{2} \|\Delta^T\|_F^2  &\leq 	4\sqrt{r}\left\| \sumt w_t {\AA^t}^* (\h^t - \z^t)  \right\|	 \|\Delta^T\|_F  + \frac{704r}{p}(\EEE( \|\bSigma_R\|  ))^2 a^2\\
&\leq 	\frac{16r}{p}\left\| \sumt w_t {\AA^t}^* (\h^t - \z^t)  \right\|^2 + 	\frac{p}{4} \|\Delta^T\|_F^2   + \frac{704r}{p}(\EEE( \|\bSigma_R\|  ))^2 a^2,
\end{align*}
which further gives
\begin{equation}
\begin{aligned}
\|\Delta^T\|_F^2  \leq \frac{64r}{p^2}\left\| \sumt w_t {\AA^t}^* (\h^t - \z^t)  \right\|^2  + 	 \frac{2816r}{p^2}(\EEE( \|\bSigma_R\|  ))^2 a^2.
\label{eq:proof_mid1}
\end{aligned}
\end{equation}
According to \cite[Lemma E.2]{xu2016dynamic}, we can bound $\EEE( \|\bSigma_R\|  )$  as
\begin{align}
\EEE(\|\bSigma_R\|) \leq C \sqrt{\frac{M \log(2N)\sumt w_t^2  }{N}}
\label{eq:proof_mid2}	
\end{align}
when provided
\begin{align*}
M \geq c N \log(2N) \frac{w_{\max}^2}{\sumt w_t^2}.	
\end{align*}
Here, both $C$ and $c$ are some numerical constants, and $w_{\max} \triangleq \max\{w_1,\cdots,w_T\}$.
Then, we are left with bounding the first term in~\eqref{eq:proof_mid1}.

Let $\Z^t$ be a Gaussian random matrix with  entries following $\NN(0,\sigma_1^2)$ and $\AA^t(\Z^t) = -\z^t$.
Define $\J \triangleq \sumt w_t {\AA^t}^* (\h^t - \z^t)$, which can be rewritten as
\begin{align*}
\J &= \sumt 	w_t {\AA^t}^* \AA^t \left(\X^T-\X^t + \Z^t \right)\\
&=\sumt 	w_t {\AA^t}^* \AA^t \left( \left(\S^T -\S^t \right)\Q^\top - \Q \left(\S^T-\S^t \right)^\top + \Z^t \right)\\
&=\sumt 	w_t {\AA^t}^* \AA^t \!\left(\! \left(\sum_{i = t+1}^T \E^i \!\right)\Q^\top \!\!-\! \Q \left(\sum_{i = t+1}^T \E^i  \right)^\top \!\!\!+\! \Z^t \right)\\
&=\sumt 	w_t {\AA^t}^* \AA^t \left( {\Y^t}^\top - \Y^t   + \Z^t \right)\\
&=\sumt 	w_t {\AA^t}^* \AA^t \left( \F^t \right),
\end{align*}
where $\Y^t \triangleq \Q \left(\sum_{i = t+1}^T \E^i  \right)^\top$ and $\F^t \triangleq {\Y^t}^\top - \Y^t   + \Z^t$. Recall that the entries of $\E^i$ and $\Z^t$ satisfy $\NN(0,\sigma_2^2)$ and $\NN(0,\sigma_1^2)$, respectively. It can be seen that the entries of $\Y^t$ are independent for a given $t$. But $\Y^t$ may be correlated for different $t\in[T]$.

The analysis below is for {\em fixed} $\Y^t$ and $\F^t$. Note that
\begin{align*}
\J &=\sumt 	w_t {\AA^t}^* \AA^t \left( \F^t \right)
    = \sumt w_t \summ F^t_{i_m j_m}\e_{i_m} \e_{j_m}^\top\\
    &= \sumt w_t \summ (Y^t_{j_m i_m} - Y^t_{i_m j_m}   + Z^t_{i_m j_m})\e_{i_m} \e_{j_m}^\top \\
    &= p \sumt \frac 1 M \summ w_t N^2(Y^t_{j_m i_m} \!-\! Y^t_{i_m j_m}   \!+\! Z^t_{i_m j_m})\e_{i_m} \e_{j_m}^\top.
\end{align*}
Construct a random matrix $\G^t\in\RRR^{N\times N}$ as
\begin{align*}
\G^t = w_t N^2(Y^t_{ji} - Y^t_{ij}   + Z^t_{ij})\e_{i} \e_{j}^\top =  w_t N^2 F^t_{ij}\e_{i} \e_{j}^\top, ~~ \text{with probability } N^{-2}.	
\end{align*}
It follows that $\J$ can be represented as a sum of independent random matrices, namely,
\begin{align*}
\J = \frac p M \sumt \summ \G_m^t,	
\end{align*}
where $\{\G_m^t\}_{m=1}^M$ are independent copies of $\G^t$.

To upper bound $\|\J\|$, we will use the uncentered matrix Bernstein inequality that is introduced in the lemma below.
\begin{Lemma}\cite{tropp2015introduction,xu2016dynamic}
\label{LEM:bernstein}
Let $\{\R_m \in \RRR^{N\times N}\}_{m=1}^M$ be a finite sequence of some independent random matrices satisfying
\begin{align*}
\|\R_m - \EEE\R_m\| \leq L, ~\forall ~m\in{M}.	
\end{align*}
Denote $\rho(\bGamma)$ as the matrix variance statistic of the sum $\bGamma = \summ \R_m$:
\begin{align*}
\rho(\bGamma) &= \max \left\{\left\|\EEE\left[(\bGamma-\EEE\bGamma)(\bGamma-\EEE\bGamma)^\top \right]\right\|,~ \left\| \EEE\left[(\bGamma-\EEE\bGamma)^\top(\bGamma-\EEE\bGamma) \right]  \right\| \right\} \\
&= \max \left\{\left\|\summ \EEE\left[(\R_m-\EEE\R_m)(\R_m-\EEE\R_m)^\top \right]   \right\|, \right.\\
&\quad\quad\quad~~\left.\left\|  \summ \EEE\left[(\R_m-\EEE\R_m)^\top(\R_m-\EEE\R_m) \right] \right\| \right\}.
\end{align*}
Then, we have
\begin{align*}
\PPP(\|\bGamma-\EEE\bGamma\| \geq d) \leq 2N \text{{\em exp}}\left(\frac{-d^2/2  }{\rho(\bGamma)+Ld/3}\right)	
\end{align*}
for any $d>0$.	
\end{Lemma}

The definition of $\G^t$ yields
\begin{align*}
\EEE\G^t  = w_t N^2 \EEE \left[ F^t_{ij}\e_{i} \e_{j}^\top\right]  = w_t N^2 \sum_{i,j} N^{-2} F^t_{ij}\e_{i} \e_{j}^\top  = w_t \F^t.	
\end{align*}
Defining $\bGamma_G \triangleq \frac 1 M \sumt \summ \G_m^t$, we have $\J = p \bGamma_G$ and $\EEE \bGamma_G = \sumt w_t \F^t$.
It follows from~\cite{xu2016dynamic} that
\begin{align*}
\rho\left(\bGamma_G \right) &\leq \frac 1 M \max\left\{\!\left\| \sumt \!\EEE\!\left[ \G^t {\G^t}^\top \!\right]\!\right\| , \left\|\sumt \!\EEE\!\left[  {\G^t}^\top \G^t \right]\!\right\|\!  \right\} \\
&= \frac{N^2}{M} \max \left\{\max_i \sumt \sum_{j=1}^N w_t^2(F^t_{ij})^2, ~ \max_j \sumt \sum_{i=1}^N w_t^2(F^t_{ij})^2  \right\} \\
& = \frac{N^2}{M} \max \left\{\max_i \alpha_i,  \max_j  \beta_j  \right\} \triangleq  \rho_0,	
\end{align*}
where we have defined
\begin{align}
\alpha_i = \sumt \sum_{j=1}^N w_t^2(F^t_{ij})^2, \quad \beta_j = \sumt \sum_{i=1}^N w_t^2(F^t_{ij})^2.	
\label{eq:def_abij}
\end{align}
Then, the remaining work is to upper bound $\|\G_m^t - \EEE\G_m^t\|$ for all $t\in[T]$, $m\in[M]$ and $\rho_0$.

\noindent
{\bf (1) Bounding $\|\G_m^t - \EEE\G_m^t\|$:} We first bound $\|\G_m^t\|$ and $\|w_t\F^t\|$. In particular, we have
\begin{align*}
\|\G_m^t\| &= w_t N^2 \|(Y^t_{j_m i_m} - Y^t_{i_m j_m}   + Z^t_{i_m j_m})\e_{i_m} \e_{j_m}^\top\|	\\
&\leq  N^2 \max_{t,i,j} w_t|Y^t_{ji} - Y^t_{ij}   + Z^t_{ij}| \\
& \leq 2N^2\left(\max_{t,i,j} w_t |Y^t_{ij}| +  \max_{t,i,j} w_t |Z^t_{ij}|  \right),
\end{align*}
and
\begin{align*}
\|w_t\F^t\| &= w_t \|\F^t\| \leq N w_t \|\F^t\|_\infty \leq N \max_{t,i,j} w_t|Y^t_{ji} - Y^t_{ij}   + Z^t_{ij}|  \\
& \leq 2N\left(\max_{t,i,j} w_t |Y^t_{ij}| +  \max_{t,i,j} w_t |Z^t_{ij}|  \right).
\end{align*}
It follows from~\cite{xu2016dynamic} that
\begin{equation}
\begin{aligned}
&\PPP\left( \max_{t,i,j} w_t |Y^t_{ij}| \!\leq\!  2\sqrt{ \log(2T N^3) \max_t w_t^2 \frac{\mu^2 r}{N} \sigma_2^2(T-t)  }  \right)\geq 1-N^{-1},\\
&\PPP\left( \max_{t,i,j} w_t |Z^t_{ij}| \leq  \sqrt{2 \log(2T N^3) \max_t w_t^2 \sigma_1^2}  \right) 	\geq 1-N^{-1}.
\label{eq:proof_prob}
\end{aligned}	
\end{equation}
With the triangle inequality, we have
\begin{equation}
\begin{aligned}
\|\G_m^t - \EEE\G_m^t\| \leq \|\G_m^t\| + \|\EEE\G_m^t\|
\leq 4N^2	\left(\max_{t,i,j} w_t |Y^t_{ij}| +  \max_{t,i,j} w_t |Z^t_{ij}|  \right).
\label{eq:proof_tria}
\end{aligned}
\end{equation}
Combining~\eqref{eq:proof_prob} and~\eqref{eq:proof_tria} yields
\begin{align*}
&\PPP\left( \|\G_m^t - \EEE\G_m^t\| \leq  CN^2\sqrt{ \log(2T N^3) }\right.\\
&\left. \left(\!\frac{\sigma_2 \mu \sqrt{r}}{\sqrt{N}}\!\sqrt{\max_t w_t^2 (T\!-\!t)} \!+\! \sigma_1\sqrt{\max_t w_t^2}\right) \! \right) \!\geq\! 1-cN^{-1},	
\end{align*}
where $C$ and $c$ denote some numerical constants. Therefore, we can set $L$ as
\begin{align*}
L \triangleq C \frac{N^2}{M}\sqrt{ \log(2T N^3) } &\left( \frac{\sigma_2 \mu \sqrt{r}}{\sqrt{N}}\sqrt{\max_t w_t^2 (T-t)}  + \sigma_1\sqrt{\max_t w_t^2} \right).	
\end{align*}

\noindent
{\bf (2) Bounding $\rho_0$:} Recall that $\rho_0 = \frac{N^2}{M} \max \left\{\max_i \alpha_i,  \max_j  \beta_j  \right\}$ with $\alpha_i$ and $\beta_j$ defined in~\eqref{eq:def_abij}.
Next, we will first bound $\max_i \alpha_i$ and $\max_j  \beta_j$ in sequence. Note that
\begin{align*}
\alpha_i &= \sumt w_t^2 \sum_{j=1}^N (Y^t_{ji} - Y^t_{ij}   + Z^t_{ij})^2 \\
&\leq 4\sumt w_t^2 \sum_{j=1}^N \left[(Y^t_{ji})^2 + (Y^t_{ij})^2   + (Z^t_{ij})^2 \right],\\
\beta_j &= \sumt w_t^2 \sum_{i=1}^N (Y^t_{ji} - Y^t_{ij}   + Z^t_{ij})^2 \\
&\leq 4\sumt w_t^2 \sum_{i=1}^N \left[(Y^t_{ji})^2 + (Y^t_{ij})^2   + (Z^t_{ij})^2 \right].	
\end{align*}
For any $i,j\in[N]$ and any $t\in[T]$, $\sum_{j=1}^N (Z^t_{ij})^2$ and $\sum_{i=1}^N (Z^t_{ij})^2$ are random variables satisfying the Chi-square distribution, namely,
\begin{align*}
\sum_{j=1}^N (Z^t_{ij})^2 \sim \sigma_1^2 \chi^2(N), \quad 	\sum_{i=1}^N (Z^t_{ij})^2 \sim \sigma_1^2 \chi^2(N).
\end{align*}
Then, we have
\begin{align*}
&\PPP\left( \max_i \sumt w_t^2 \sum_{j=1}^N (Z^t_{ij})^2 \leq C N \sumt w_t^2 \sigma_1^2     \right)	 \geq 1-TN \text{exp}(-N),\\
&\PPP\left( \max_j \sumt w_t^2 \sum_{i=1}^N (Z^t_{ij})^2 \leq C N \sumt w_t^2 \sigma_1^2     \right)	 \geq 1-TN \text{exp}(-N),
\end{align*}
by applying the tail bound of Chi-squared variable and the standard union bound~\cite{xu2016dynamic}. Similarly, we can also get\footnote{One can refer to \cite{xu2016dynamic} for more details.}
\begin{align*}
&\PPP\left( \max_i \sumt w_t^2 \sum_{j=1}^N (Y^t_{ij})^2 \leq C \mu^2r \sumt(T-t) w_t^2 \sigma_2^2     \right) \geq 1-TN \text{exp}(-N),\\
&\PPP\left( \max_j \sumt w_t^2 \sum_{i=1}^N (Y^t_{ij})^2 \leq C N\sumt(T-t) w_t^2 \sigma_2^2     \right)	  \geq 1-TN \text{exp}(-N).	
\end{align*}
Note that
\begin{align*}
&\max_i \sumt w_t^2 \sum_{j=1}^N (Y_{ji}^t)^2 =  \max_j \sumt w_t^2 \sum_{i=1}^N (Y_{ij}^t)^2,\\
&\max_j \sumt w_t^2 \sum_{i=1}^N (Y_{ji}^t)^2 =  \max_i \sumt w_t^2 \sum_{j=1}^N (Y_{ij}^t)^2.	
\end{align*}
Then, we can get
\begin{align*}
&\PPP\left( \max_i \alpha_i \leq C N\sumt w_t^2 \left(  \sigma_1^2 +  (1 + \frac{\mu^2r}{N})  (T-t)   \sigma_2^2 \right)       \right)	  \geq 1-cTN \text{exp}(-N),\\
&\PPP\left( \max_j \beta_j \leq C N\sumt w_t^2 \left(  \sigma_1^2 +  (1 + \frac{\mu^2r}{N})  (T-t)   \sigma_2^2 \right)      \right)  \geq 1-cTN \text{exp}(-N),	
\end{align*}
which further gives
\begin{align*}
&\PPP\left( \rho_0 \leq C \frac{N^3}{M} \sumt w_t^2 \left(  \sigma_1^2 +  (1 + \frac{\mu^2r}{N})  (T-t)   \sigma_2^2 \right) \triangleq \nu       \right)	  \geq 1-cTN \text{exp}(-N).	
\end{align*}
Now, we are ready to apply Lemma~\ref{LEM:bernstein} and thus get
\begin{align*}
\PPP\left(\left\|\frac 1 M \sumt \summ \G_m^t - \sumt w_t\F^t     \right\| \geq d \right)  \leq 2N \text{exp}\left( \frac{-d^2/2 }{\nu+Ld/3  }\right).	
\end{align*}
Letting $d = 2\sqrt{\log(2N)\nu}$, we further obtain
\begin{align*}
\PPP\left(\!\left\|\! \frac 1 M \!\sumt \!\summ \!\G_m^t \!-\! \sumt\! w_t\F^t     \right\| \!\geq\!  2\sqrt{\log(2N)\nu} \right) \!\leq \!\frac 1 2 N^{-1}
\end{align*}
if $\nu$ dominates the denominator of the exponential term, i.e.,
\begin{align*}
	\nu \geq \frac 1 3 L d  = \frac 2 3 L  \sqrt{\log(2N)\nu},
\end{align*}
which can be satisfied if
\begin{align*}
M &\geq  CN\log(2TN^3)\log(2N) \frac{ \left( \frac{\sigma_2 \mu \sqrt{r}}{\sqrt{N}}\sqrt{\max_t w_t^2 (T-t)} + \sigma_1\sqrt{\max_t w_t^2} \right)^2}{ \sumt w_t^2 \left(  \sigma_1^2 +  (1 + \frac{\mu^2r}{N})  (T-t)   \sigma_2^2 \right)}\\
&\geq   CN\log(2TN^3)\log(2N) \frac{ \left( \frac{\sigma_2 \mu \sqrt{r}}{\sqrt{N}}\sqrt{\max_t w_t^2 (T-t)} + \sigma_1\sqrt{\max_t w_t^2} \right)^2}{ \sumt w_t^2 \left(  \sigma_1^2 +  2  (T-t)   \sigma_2^2 \right)},
\end{align*}
where the second inequality follows from $\frac{\mu^2r}{N} \leq 1$.

The remaining work is to bound $\left\| \sumt w_t\F^t \right\|$. Recall that $\F^t = {\Y^t}^\top - \Y^t   + \Z^t$. Note that each entry of $\F^t$ is a Gaussian random variable with variance not greater than $2\sigma_2^2(T-t)+\sigma_1^2$. Then, we have
\begin{align*}
&\PPP\left(\left\| \sumt w_t\F^t  \right\| \leq 2\sqrt{N\sumt w_t^2 (2\sigma_2^2(T-t)+\sigma_1^2) }   \right)	 \geq 1-C_1 \text{exp}(-C_2 N),
\end{align*}
where $C_1$, $C_2$ are  numerical constants. Note that $C_1 \text{exp}(-C_2 N) \ll TN \text{exp}(-N)$. Finally, we can bound $\|\J\|^2$ as
\begin{align*}
&\|\J\|^2  = p^2 \left\|\frac 1 M \sumt \summ \G_m^t	\right\|^2\\
\leq &p^2 \left( \left\| \frac 1 M \sumt \summ \G_m^t - \sumt w_t\F^t     \right\| +  \left\| \sumt w_t\F^t  \right\|       \right)^2\\
\leq &  p^2 \!\left(\! 2\sqrt{\log(2N)\nu} +  2\sqrt{N\sumt w_t^2 (2\sigma_2^2(T-t)+\sigma_1^2) }       \right)^2\\
\stack{\ding{172}}{\leq} &Cp^2\!\!\left(\frac{N^3}{M}\log(2N) \sumt w_t^2 \left(  \sigma_1^2 +  (1 + \frac{\mu^2r}{N})  (T-t)   \sigma_2^2 \right)   +  N\sumt w_t^2 (2\sigma_2^2(T-t)+\sigma_1^2)  \!\!  \right)\\
\stack{\ding{173}}{\leq} & Cp^2 N \max\left\{N^2\frac{\log(2N) }{M}, 1     \right\}\sumt w_t^2 (2\sigma_2^2(T-t)+\sigma_1^2)\\
\stack{\ding{174}}{\leq}  &C\frac{M }{N}\log(2N)\sumt w_t^2 (2\sigma_2^2(T-t)+\sigma_1^2)
\end{align*}
with probability at least $1 - c_1N^{-1} - c_2TN\text{exp}(-N)$.
Here, \ding{172} follows from the inequality $(a+b)^2 \leq 2(a^2 + b^2)$. \ding{173} follows from $\frac{\mu^2r}{N} \leq 1$. \ding{174} follows from $M \leq N^2\log(2N)$ and $p = \frac{M}{N^2}$.
Together with the two inequalities in~\eqref{eq:proof_mid1} and~\eqref{eq:proof_mid2}, we finish the proof of~\cref{THM_error}.

\end{document}